\documentclass[apj,twocolumn]{openjournal}
\usepackage{graphicx}
\usepackage{xcolor}
\usepackage{amsmath}
\usepackage{amsfonts}
\usepackage{amssymb}
\usepackage{upgreek}
\usepackage[pdfborder={0 0 0}]{hyperref}
\usepackage{txfonts}
\usepackage{lipsum} 
\newcommand{\orcidauthor}[3]{\author{\href{http://orcid.org/#1}{#2 \openin1 Orcid-ID.png \ifeof1 \else \hskip2pt\includegraphics[width=9pt]{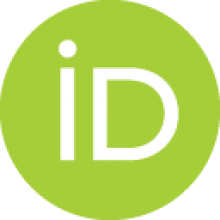}\fi}$^{#3}$}}

\begin{document}

\title{A Square Kilometre Array Pulsar Census\vspace{-20pt}}


\orcidauthor{0000-0002-4553-655X}{E.~F.~Keane}{1}
\orcidauthor{0000-0002-6558-1681}{V.~Graber}{2}
\orcidauthor{0000-0002-2034-2986}{L.~Levin}{3}
\orcidauthor{0000-0001-7509-0117}{C.~M.~Tan}{4}
\orcidauthor{0000-0002-5927-0481}{O.~A.~Johnson}{1}
\orcidauthor{0000-0002-3616-5160}{C.~Ng}{5}
\orcidauthor{0000-0002-8118-255X}{C.~Pardo-Araujo}{6,7}
\orcidauthor{0000-0003-2781-9107}{M.~Ronchi}{8}
\orcidauthor{0000-0003-1779-4532}{D.~Vohl}{8}
\orcidauthor{0000-0001-8018-1830}{M.~Xue}{9}
\author{The SKA Pulsar Science Working Group}

\affiliation{$^1$ School of Physics, Trinity College Dublin, College Green, Dublin 2, D02 PN40, Ireland}
\affiliation{$^2$ Department of Physics, Royal Holloway, University of London, Egham, TW20 0EX, UK}
\affiliation{$^3$ Jodrell Bank Centre for Astrophysics, University of Manchester,
Oxford Road, M13 9PL, Manchester, UK}
\affiliation{$^4$ International Centre for Radio Astronomy Research (ICRAR), Curtin University, Bentley, WA 6102, Australia}
\affiliation{$^5$ Laboratoire de Physique et Chimie de l'Environnement et de l'Espace - Universit\'e d'Orl\'eans/CNRS, 45071, Orl\'eans Cedex 02, France}
\affiliation{$^6$ Institute of Space Sciences (CSIC-ICE), Campus UAB, Carrer de Can Magrans s/n, 08193, Barcelona, Spain}
\affiliation{$^7$ Institut d’Estudis Espacials de Catalunya (IEEC), Carrer Gran Capità 2–4, 08034 Barcelona, Spain}
\affiliation{$^8$ ASTRON, the Netherlands Institute for Radio Astronomy, Oude Hoogeveensedijk 4, 7991 PD Dwingeloo, The Netherlands}
\affiliation{$^9$ National Astronomical Observatories, Chinese Academy of Sciences, Beijing 100101, China}


\begin{abstract}
Most of the pulsar science case with the Square Kilometre Array (SKA) depends on long-term precision pulsar timing of a large number of pulsars, as well as astrometric measurements of these using very long baseline interferometry (VLBI). But before we can time them, or VLBI them, we must first find them. Here, we describe the considerations and strategies one needs to account for when planning an all-sky blind pulsar survey using the SKA. Based on our understanding of the pulsar population, the performance of the now-under-construction SKA elements, and practical constraints such as evading radio frequency interference, we project pulsar survey yields using two complementary methods for a number of illustrative survey designs, combining SKA1-Low and SKA1-Mid Bands 1 and 2 in a variety of ways. A composite survey using both Mid and Low is optimal, with Mid Band 2 focused in the plane. We find that, given its much higher effective area and survey speed, the best strategy is to use SKA1-Low to cover as much sky as possible, ideally also overlapping with the areas covered by Mid. 
%
In our most realistic scenario, we find that an all-sky blind survey with Phase 1 of the SKA with the AA* array assembly will detect $\sim10,000$ slow pulsars and $\sim 800$ millisecond pulsars (MSPs) if SKA1-Mid covers the region within $5\deg$ of the plane, while higher latitudes will be covered with SKA1-Low. The yield with AA4 is $\sim 20\%$ higher. One could increase these numbers by increasing the range covered by SKA1-Mid Bands 1 and 2, at the cost of a considerably longer survey. The pulsar census will enable us to set new constraints on the uncertain physical properties of the entire neutron star population. This will be crucial for addressing major SKA science questions including the dense-matter equation of state, strong-field gravity tests, and gravitational wave astronomy.
\end{abstract}
\maketitle
\vspace{-2\baselineskip}


\section{Introduction}

Pulsar science is one of the foremost drivers for the Square Kilometre Array (SKA) Observatory~\citep{bbg+15}. The possibilities in pulsar science highlighted in the 2015 SKA Science Case~\citep{ks15} have, in the intervening decade, motivated the design of many components of the SKA, which is now under construction in South Africa and Australia. In this paper, we revise our earlier estimates~\citep{kbk+15,lab+18} for the number of pulsars detectable by SKA Phase 1 (hereafter SKA1) using a variety of blind pulsar survey approaches. Updated estimates are timely for a number of reasons. Firstly, immediately after the publication of our 2015 estimates, SKA1 was `re-baselined'; the effect was a nominal loss of $50\%$ and $30\%$ in collecting area for SKA1-Low and Mid, respectively. In addition to this, the design went from incompatible `desirements' of diverse user groups, to a rigorously tested set of requirements determined via a systems engineering approach~\citep{L1v12}. Many additional design choices and optimisations have occurred in the last decade. As well as the parameters of the SKA changing, our understanding of the pulsar population has improved in many ways, often driven by new discoveries and measurements performed by SKA precursor and pathfinder instruments. Here, we factor in these changes to re-estimate what the pulsar survey yield might be from SKA1 using a two-pronged approach, discuss underlying challenges in deriving these estimates and highlight key areas that we need to address to advance our pulsar science cases.

Below, in \S~\ref{sec:pop_sims}, we describe our two population synthesis analyses, both the snapshot and evolutionary approaches. This is followed in \S~\ref{sec:ska} by our analysis of the sensitivity and survey speed of both the Mid and Low arrays. As subsets of the full arrays can be combined, in what is termed a `sub-array', there are a number of sensitivity and field-of-view combinations possible so our analysis is performed as a function of increasing sub-array size. With our calculated and chosen survey parameters, we then present our population synthesis results and our expected survey yields when employing each of three options for composite surveys using Low, Mid Band 1 and Mid Band 2 in tandem in different ways. Finally in \S~\ref{sec:discussion}, we discuss our estimates, and the magnitude and reasons for systematic uncertainties in these. We conclude with a `wish list' of activities that can already be undertaken to improve pulsar population modeling and survey yield estimates.


\section{Population Synthesis}\label{sec:pop_sims}
Pulsar population synthesis is an attempt to determine the intrinsic underlying population of pulsars in the Galaxy~\citep{fk06,lfl+06,khbm08,gmvp14,jk17}, along with their key properties, based on the known observed population and our understanding of observational selection effects and other biases, as well as on our theoretical understanding of neutron star astrophysics which cannot be derived from the observations alone. Population synthesis can be undertaken in a number of ways. Here, we perform the modeling in two ways, firstly using a `snapshot' approach~\citep{lfl+06,blrs14} and secondly using an evolutionary method~\citep{grpn24, prgr25}. While the former method models the Galactic pulsar population based on currently observed properties, the latter evolves pulsars based on a set of initial conditions to reach the current population. Subsequent observational filters applied (see \S~\ref{sec:ska}) are, however, identical.


\subsection{Snapshot} \label{sec:snapshot}

\begin{figure}[b]
   \centering
   \includegraphics[trim = {10mm 20mm 10mm 20mm}, clip, width=\hsize,]{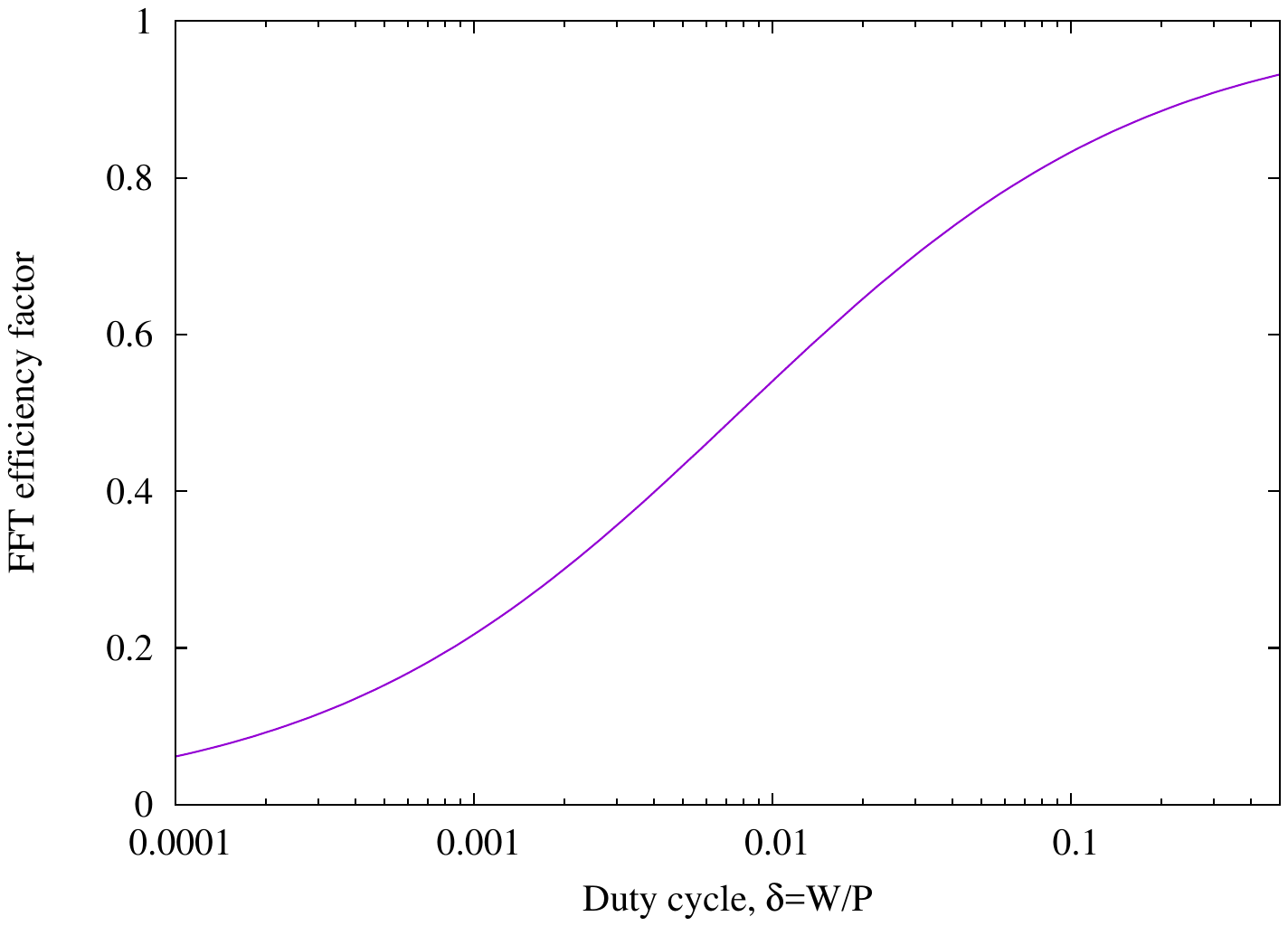}
   \caption{Shown is the FFT correction factor as a function of duty cycle, as determined by \citet{mbs+20}; this is the response when $32$ harmonics are summed. The efficiency is relative to perfectly phase-coherent signal-to-noise estimates, which the fast folding algorithm~\citep{mbs+23} gets closer to.}
   \label{fig:fft_correction}
\end{figure}

The snapshot pulsar population synthesis was undertaken using \textsc{psrpoppy}~\citep{blrs14}, which is a translated and expanded version of its predecessor \textsc{psrpop}~\citep{lfl+06}. This population synthesis approach tries to reproduce the current observed distributions of pulsar properties, and with this achieved, the resulting population can be used to make predictions for any given survey parameters. We first updated the software\footnote{\texttt{https://github.com/evanocathain/PsrPopPy3}} to correct a number of issues with inconsistent survey parameters.
The most substantive change is an update to the signal-to-noise ratio calculation. Pulsars are predominantly found via Fourier-domain methods. As such, we included the duty cycle-dependent efficiency factor determined by \citet{mbs+20} as appropriate for incoherent searches based on Fast Fourier Transforms. Figure~\ref{fig:fft_correction} shows this correction factor. The overall effect is to suppress narrow duty cycle pulsar signals. As the duty cycles in the long-period (hereafter `slow') pulsar population are typically narrow in general, whereas those of the faster spinning millisecond pulsar population are wider, the deleterious effect is more pronounced in the slow pulsar population.

As a result of these changes, and the slight increase in pulsar detections in surveys previously considered to be complete, we re-performed the analysis of \citet{blv13} to determine the spectral index distribution that is consistent with three archival Parkes surveys. These were namely the Parkes Southern Pulsar Survey~\citep{mld+96,lml+98} at $70$~cm, the Parkes Multi-beam Pulsar Survey~\citep{mlc+01,mhl+02,kbm+03,hfs+04,fsk+04,lfl+06,ekl09,kel+09,kle+10,emk+10,kek+13} at $20$~cm, and the Methanol Multi-Beam (MMB) survey~\citep{ojk+08,bjl+11} at $6.5$~GHz. The number of slow (total) pulsars found to date in these surveys, which is used for calibration below, are: 279 (298), 1115 (1158) and 18 (18). We choose these surveys as we consider them to be the most complete. Many surveys performed since, are still being successfully mined for new discoveries (e.g.~\citealt{mlk+24,old+25}). Further, we wished to re-do the analysis of \citet{blv13}, which used these three surveys, keeping changes to the minimum. As such, we did not modify parameters such as the current-day Galactic exponential scale height, taken to be $330$~pc for slow pulsars and $500$~pc for millisecond pulsars, and we used the luminosity distribution from \citet{fk06}. These latter choices are somewhat standard in the snapshot approach~\citep{rl10,xzx+23} and in an effort to make an apples-with-apples comparison to our previous estimates~\citep{kbk+15,lab+18} we keep these. However, a wider effort is ongoing to explore these and other parameters in tandem with spectral and other properties. For now, we simply note that an increased scale height, as implied by some studies~\citep{h25}, would result in a \textit{higher} pulsar yield for SKA1-Low. We also note that in the snapshot method we are sampling from period and luminosity distributions and, by definition, do not evolve our population. As such no relation between the radio luminosity and the spin-down energy loss rate $\dot{E}_{\rm rot}$ is assumed or used in this method.

\begin{figure}
    \centering
    \includegraphics[trim = {0mm 20mm 0mm 20mm}, clip, width=\hsize,]{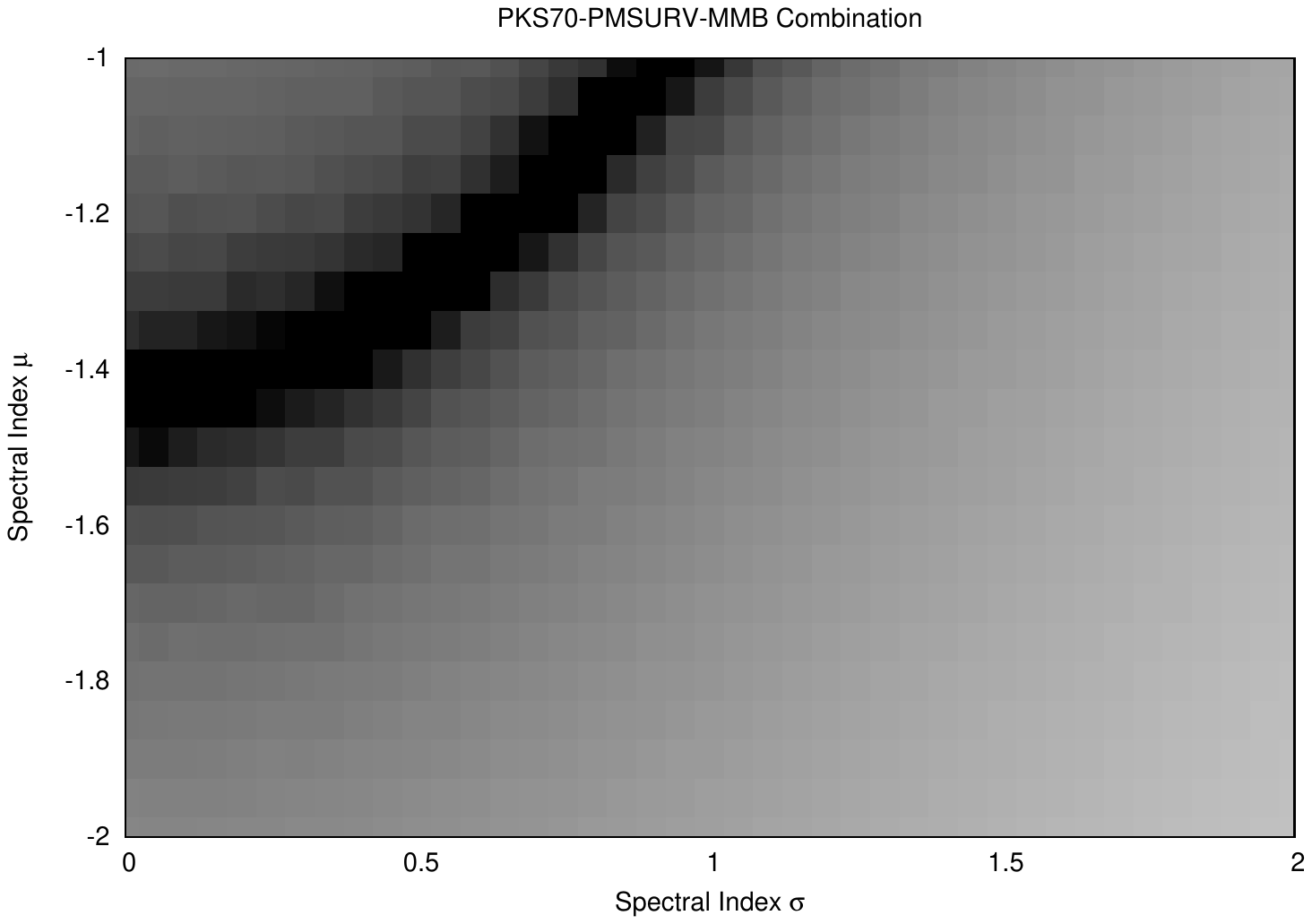}
    \includegraphics[trim = {0mm 0mm 0mm 20mm}, clip,width=\hsize,]{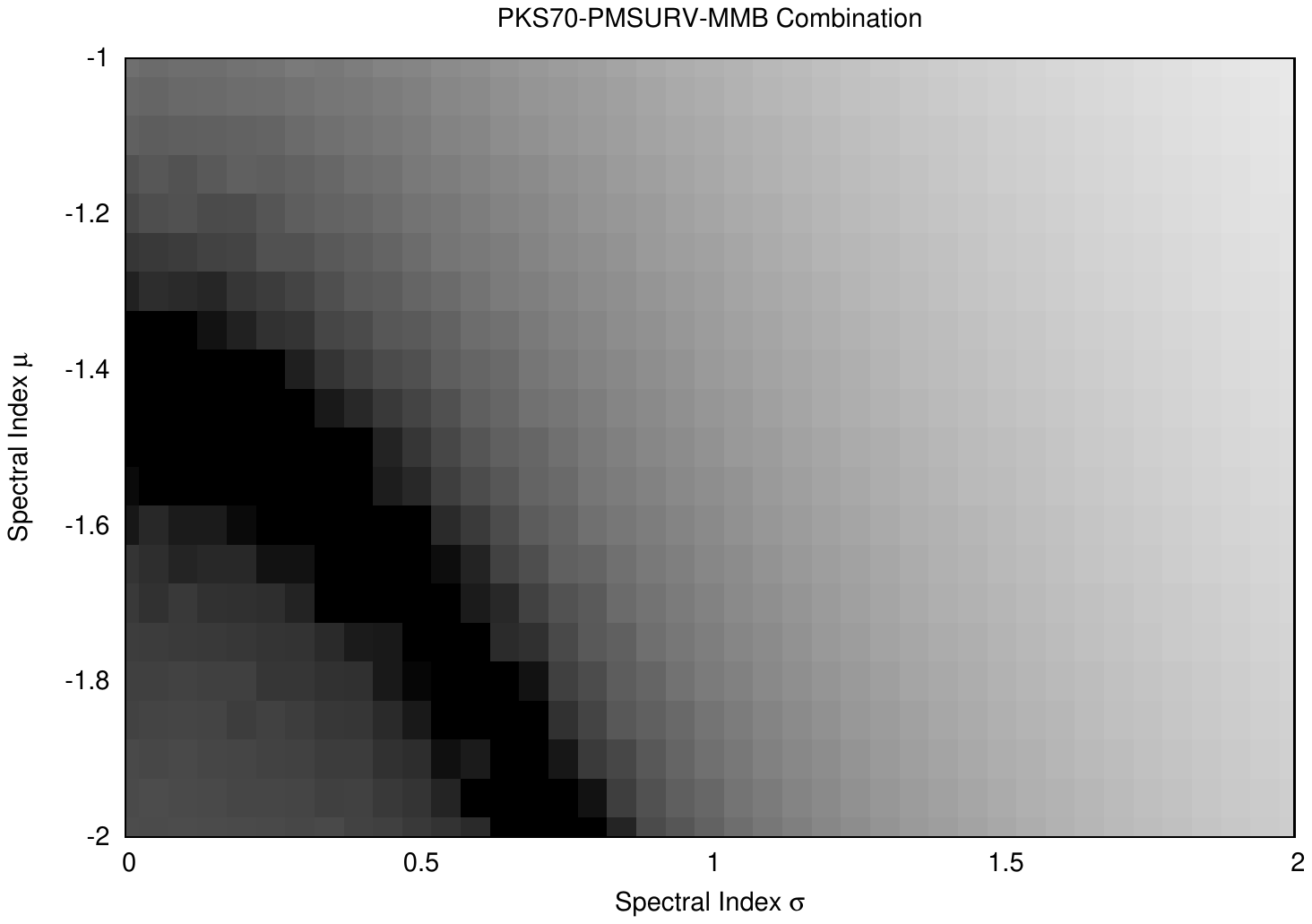}
    \includegraphics[trim = {0mm 20mm 0mm 20mm}, clip,width=\hsize,]{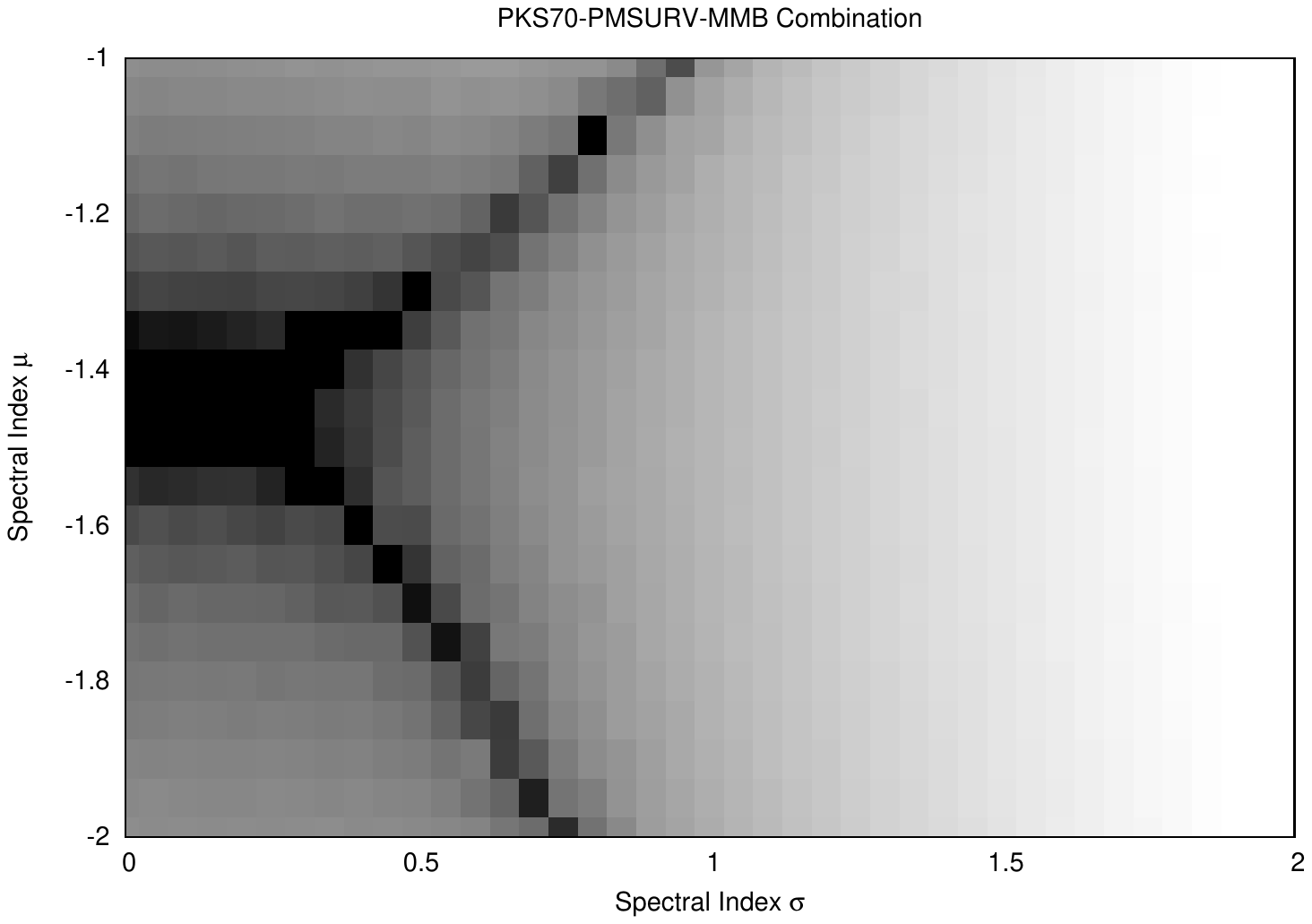}
    \caption{The top panel shows the confidence intervals, on a logarithmic colour scale, for the simulated Parkes $70$~cm survey yields. The black curve can be considered as the range of spectral index parameters, where both the $20$~cm and $70$~cm yields match acceptably. The top left region of the parameter space produces too few pulsars, the bottom right produces too many. The middle panel shows the corresponding information for the combination of the $20$~cm and $6.5$~GHz surveys. Here, the bottom left produces too few pulsars, the top right too many. The bottom panel shows the logarithm of the product of the confidence intervals, demonstrating that spectral index values of $\mu=-1.45 \pm 0.05$ and $\sigma=0.15 \pm 0.15$ match the observed detections best.}
\label{fig:spectra}
\end{figure}

Following \citet{blv13}, we modeled the spectral index distribution as a normal distribution and sampled the two-dimensional parameter space corresponding to the mean, $\mu$, and standard deviation, $\sigma$, of this distribution. The results of this analysis that best matched the actual yields of our three pulsar surveys outlined above are shown in Figure~\ref{fig:spectra}. The final answer, based on $40$ iterations, is $\mu=-1.45 \pm 0.05$ and $\sigma=0.15 \pm 0.15$. This differs from the previous determination of $-1.41 \pm 0.06$ and $1.0 \pm 0.05$; while our mean value is consistent, our distribution is much narrower. With a narrower distribution there are less pulsars both with much flatter and much steeper spectra than the mean value, meaning lower yields for both high and low frequencies, relative to intermediate frequencies in the $1-2$~GHz range. It also differs from the \textit{observed} spectral index distribution in the known population~\citep{jsk+15,pkj+23}, which is heterogeneous in terms of spectral selection effects across the various surveys wherein these pulsars were discovered. We note that, conversely, it agrees well with the observed MSP spectral distribution determined by \citet{al22}. In what follows, we adopt our newly computed values, and the assumptions mentioned above regarding the Galactic scale height and luminosity distribution, for making the snapshot yield estimates for the SKA1 arrays.

\subsection{Evolving population}\label{sec:pop_evl}

\begin{figure*}[t!]
   \centering
   \includegraphics[trim = {0mm 0mm 0mm 0mm}, clip, width=\textwidth]{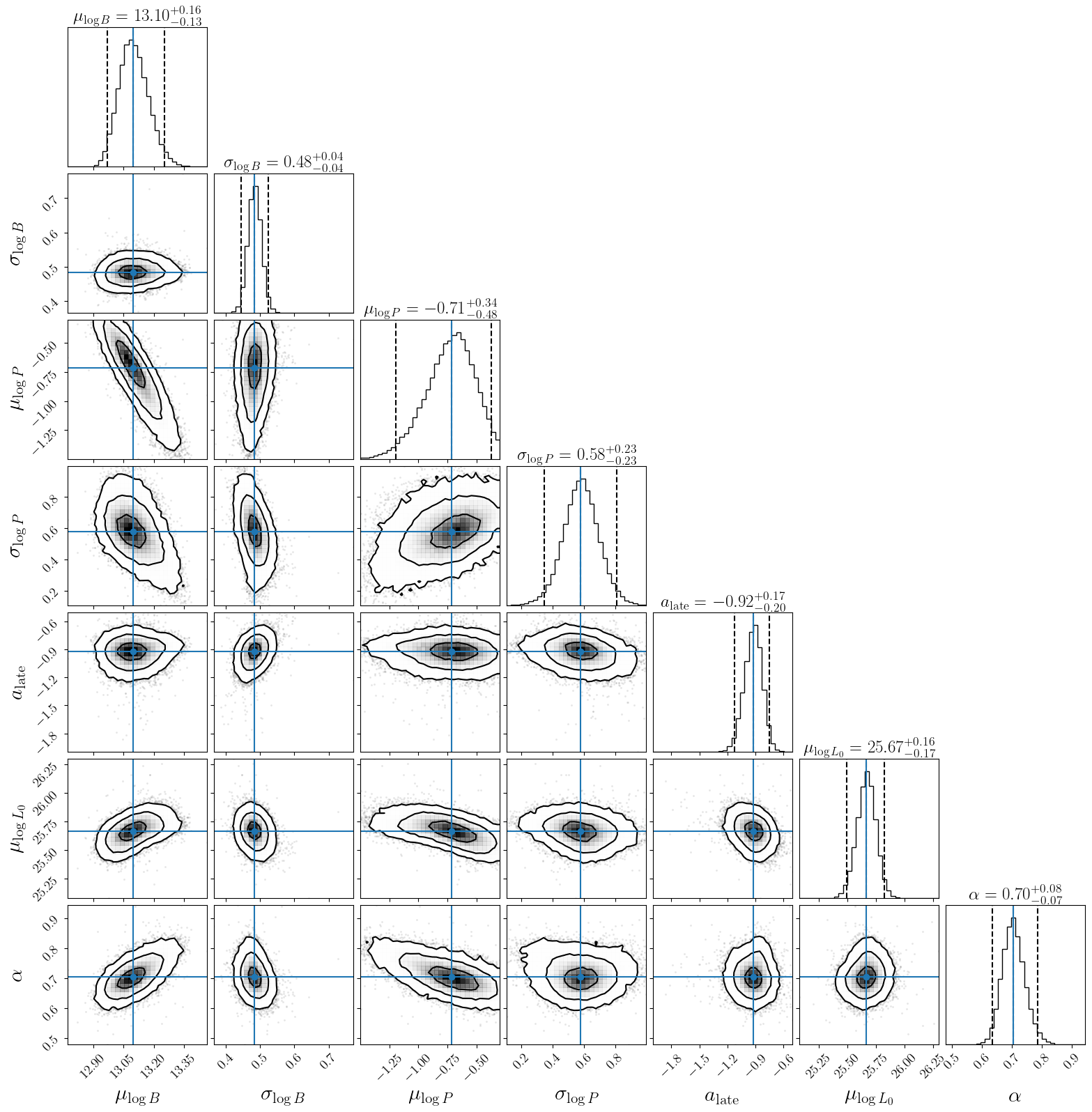}
   \caption{Corner plot with the one- and two-dimensional posterior distributions for five magneto-rotational parameters and two parameters related to the pulsars' intrinsic luminosity. We highlight the medians in light blue and show corresponding values and 95\% credible intervals above the panels. These results were obtained using the methodology outlined in \citet{prgr25} and correspond to a converged run of a sequential simulation-based inference experiment. The key difference is that the above posteriors were obtained with a different spectral index distribution (see text for details).}
   \label{fig:cornerplot}
\end{figure*}

The evolutionary population synthesis framework used in the following is based on the recent works of \citet{grpn24} and \citet{prgr25}, which use machine learning to infer a range of neutron star parameters. To model the spatial distribution of pulsars at birth, we sample positions in the Galactic plane using the electron density model from \citet{ymw17}, and adopt a vertical distribution following an exponential profile with a scale height of $180\,$pc \citep{gmvp14}. Birth velocities are drawn from the Maxwellian distribution proposed by \citet{hllk05}, with a velocity dispersion of $265\,$km$/$s. Birth spin periods and dipole magnetic field strengths are sampled from log-normal distributions with means and standard deviations allowed to vary. For the magnetic field evolution, we follow \citet{grpn24} and incorporate magneto-thermal simulations at early times and a power-law decay for pulsars older than $\sim 1$Myr, where the power-law index is also a free parameter.

We obtain the kinetic properties of each neutron star at the current time by solving the Newtonian equations of motion in the Galactic potential, while the magneto-rotational properties are evolved using a set of coupled differential equations following \citet{ptl14}. Intrinsic radio luminosities are modeled using the luminosity law from \citet{prgr25}, which scales as a power law of the rotational energy loss, introducing a further two free parameters. Beaming geometry and pulse propagation effects are also implemented as outlined in that study.

To calibrate the model, we compare synthetic populations with observed detections in three Parkes surveys: the Parkes Multibeam Pulsar Survey (1045 pulsars; \citealt{mlc+01, lfl+06}), the Swinburne Intermediate-latitude Pulsar Survey (218 pulsars; \citealt{ebsb01,jbo+09}), and the low- and mid-latitude High Time Resolution Universe survey (1095 pulsars; \citealt{kjs+10}). Note that the first count differs slightly from the value outlined in \S~\ref{sec:snapshot}, because we also apply a period derivative cut-off for our evolutionary population synthesis. Additionally, for a sub-population of these three surveys (see \citet{prgr25} for details), we incorporate flux measurements from MeerKAT as reported by \citet{pkj+23}, providing further observational constraints. The use of surveys across a wide variety of Galactic latitudes is particularly important for the evolutionary approach to probe the properties of older sources far away from the Galactic plane.

To determine the best-fit model parameters, we represent the resulting synthetic populations as density maps in the period–period derivative plane and apply a sequential simulation-based inference algorithm \citep{dgm22} to determine posterior distributions for the seven free parameters related to magneto-rotational evolution and intrinsic luminosity distribution. While results from \citet{prgr25} assumed a normal spectral index distribution with $\mu =-1.8$ following \citet{pkj+23} and $\sigma=0$, we have repeated the inference based on the parameters deduced from Figure~\ref{fig:spectra} to streamline our assumptions with those outlined in the snapshot population synthesis approach.

The resulting corner plot with one- and two-dimensional marginalised
posteriors is shown in Figure~\ref{fig:cornerplot}. The inferred
pulsar properties remain largely consistent with \citet{prgr25}
including the birth rate which ranges between $\sim 1.7 - 2.0$ neutron
stars stars per century, though the updated analysis yields a smaller
scaling factor in the luminosity law. This is expected, as the flatter
spectral index implies intrinsically brighter sources for a given
period derivative, making them more easily detectable. Note, however,
that the best inferred value for the power-law index connecting the
intrinsic luminosity and the rotational energy loss, dictating $L
\propto \dot{E}_{\rm rot}^{0.7}$, is consistent with the result of
\citet{prgr25}. Using the inferred best-fit parameters summarised in
Figure~\ref{fig:cornerplot} and a birth rate of $2$ neutron stars per
century with stars evolved up to a maximum of $2\times10^9\,$yrs to
match the observed detection counts outlined above, we then compute a
complementary set of yield estimates for the SKA1 arrays within the
evolutionary population synthesis framework.


\section{SKA Parameters}\label{sec:ska}

\begin{figure} 
    \centering
    \includegraphics[trim = {10mm 20mm 10mm 20mm}, clip, width=\hsize]{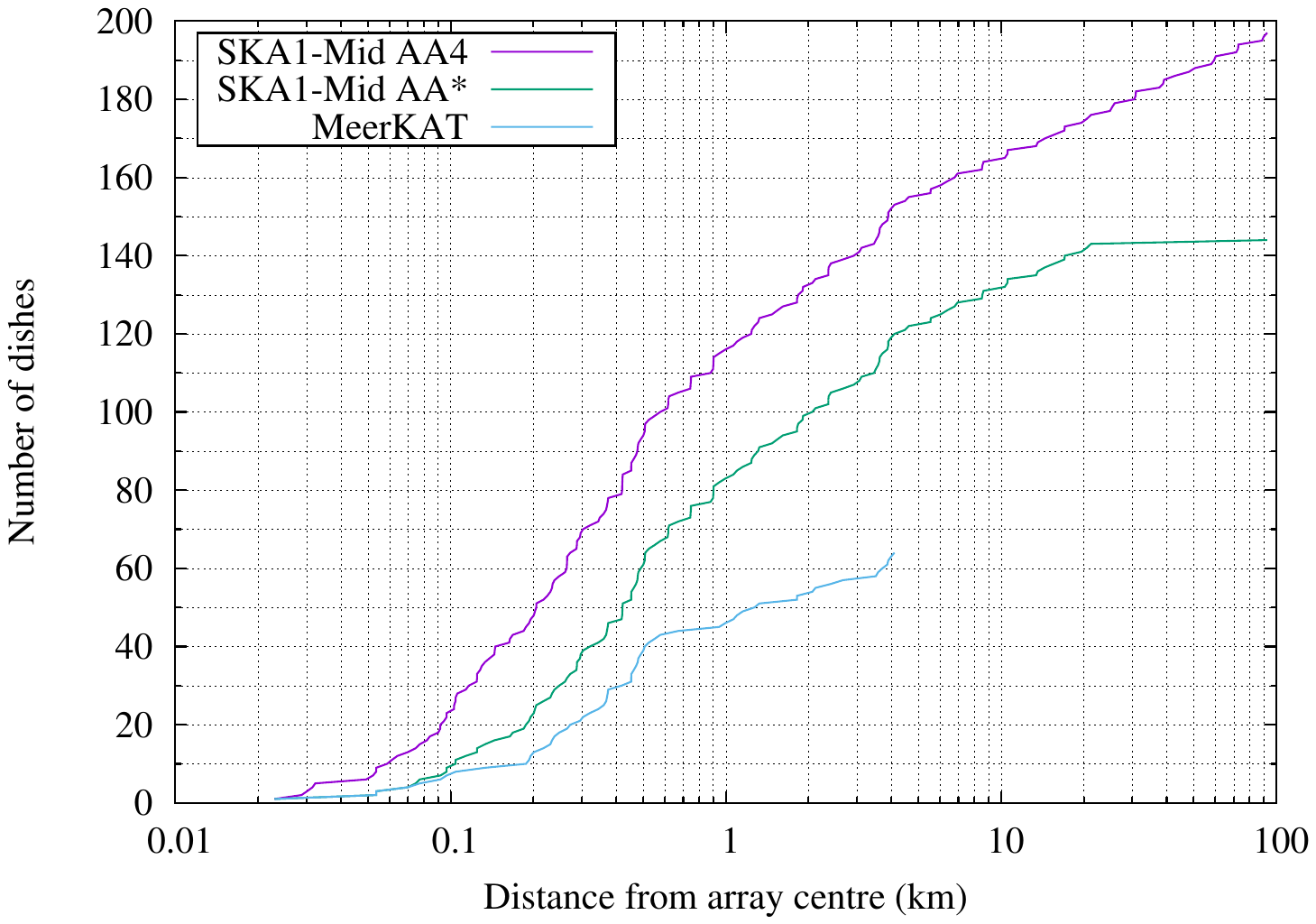}
    \includegraphics[trim = {10mm 20mm 10mm 20mm}, clip, width=\hsize]{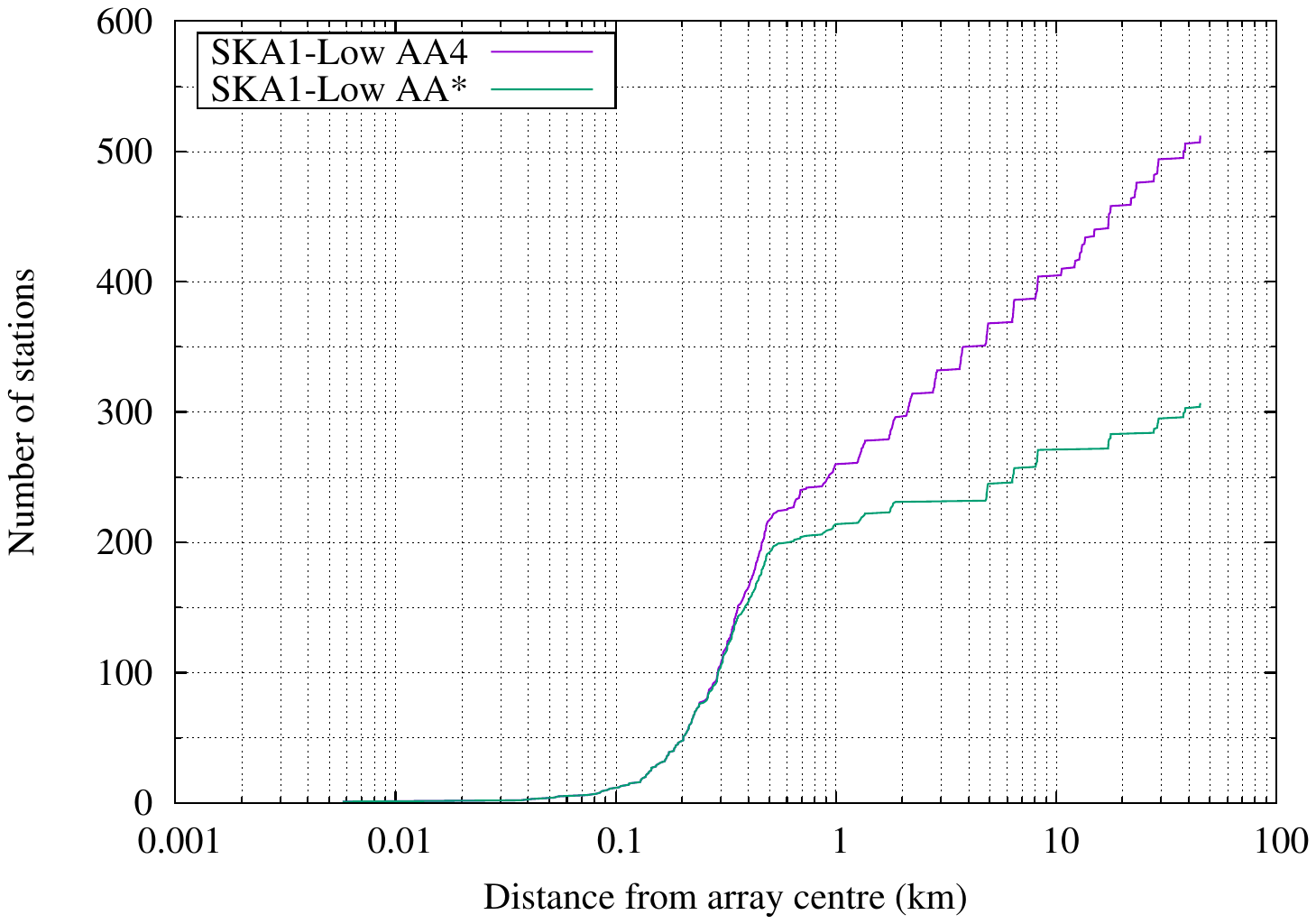}
    \caption{Shown are the cumulative number of elements for the SKA1-Mid and SKA1-Low arrays as a function of distance from the array centres; dishes for SKA1-Mid in the top panel and stations for SKA1-Low in the bottom panel. Both figures show the comparison between AA* (purple) and AA4 (green). For SKA1-Mid, the MeerKAT array standalone (blue) is also shown for comparison.}
    \label{fig:elements_versus_radius}
\end{figure}

The SKA is modular by definition and is being built as such\footnote{\scriptsize{\texttt{https://www.skao.int/en/science-users/599/scientific-timeline}}}. Various check-points along the way are being released as completed milestones; these are called array assemblies. Array assembly 0.5 (AA0.5) is followed by AA1, AA2, AA* and AA4. The first pulsar observations and first images\footnote{\scriptsize{\texttt{https://www.skao.int/en/news/621/ska-low-first-glimpse-universe}}} have already been obtained in the past year with SKA1-Low AA0.5. Here, we will consider the pulsar yield for the latter two array assemblies---AA* and AA4. Figure~\ref{fig:elements_versus_radius} shows the distribution of elements for both SKA1-Mid and SKA1-Low between AA* and AA4. For a radio array, one can combine any number of elements depending on what is desired for the observation. Coherently combining an array of $N$ elements results in a sensitivity, which is $N\times$ better than a single element. However, the field of view of such a tied-array beam, scales as $1/D^2$, where $D$ is the maximum element separation. For a survey, one therefore has a choice on how to balance sensitivity (depth of survey) and field of view (breadth of survey). Consequently, SKA1 pulsar survey parameters are informed by a number of factors, which we elaborate upon now; we do so separately for the Mid and Low arrays. In what follows, we will see that it is the core of the arrays that is most crucial for untargeted pulsar searches. For targeted pulsar searches, as e.g. performed of globular clusters~\citep{Bagchi2025_SKA_GlobClust} or the Galactic Centre~\citep{Abbate2025_SKA_GalCen}, as well as for pulsar timing observations~\citep{Shannon2025_SKA_SKAPTA}, this is not the case. In those scenarios, one can phase up the full array as the field of view is not important. Comparing the configurations AA* and AA4 in Figure~\ref{fig:elements_versus_radius}, it is evident that the SKA1-Mid array is initially more `hollowed-out' than SKA1-Low as AA4 will have more dishes in the inner $1$~km region. For Low, AA* and AA4 are comparable in the inner kilometre. We now discuss the relevant survey parameters for SKA1-Mid and SKA1-Low.

\subsection{SKA1-Mid Survey Parameters}

Many of the relevant parameters for a pulsar search are fixed and are part of the SKA design~\citep{L1v12}, but some vital parameters are configurable so that one has a choice in survey design. Some of the key fixed parameters\footnote{We note that the Pulsar System for Timing (PST) also has some searching capabilities but is limited to $16$ tied-array beams; in this case voltage beams are available. The PST might therefore be preferable for deep offline targeted searches of pre-identified targets.} of the Pulsar System for Searching (PSS) are shown in Table~\ref{tab:PSS_specs}. We note that the PSS can perform real-time Fourier domain acceleration searches on pointings as long as $10$~min, and real-time single pulse searches on pointings of up to $30$~min, on $1500$ ($1125$) tied-array Stokes $I$ beams in AA4 (AA*). Thus, in our modelling, we choose $10$~min pointings. The maximum acceleration range is $\pm 350\;\mathrm{m}\,\mathrm{s}^{-2}$ for a $500$~Hz pulsar, higher for slower pulsars. If performing a full acceleration search one can perform $500$ user-specified dispersion measure trials, and a much larger number of non-accelerated dispersion measure trials. We note that while we expect more than $300$~MHz of bandwidth to be eventually available (priv. comm., Karastergiou \& de Selby) we choose to use the current widely communicated specification of $300$~MHz in our calculations.

\begin{table}
    \caption{Pulsar search specifications in the PSS systems and their values for SKA1-Low and SKA1-Mid.}
    \label{tab:PSS_specs}
    \centering
    \begin{tabular}{lcc}
    Parameter & Low Value & Mid Value \\
    \hline
    Number tied-array Stokes $I$ beams (AA*) & 250 & 1125 \\
    Number tied-array Stokes $I$ beams (AA4) & 500 & 1500 \\
    Instantaneous bandwidth per beam & $100$~MHz & $300$~MHz \\
    Number of frequency channels & 8192 & 4096 \\
    Frequency resolution & 13~kHz & 73~kHz \\
    Max. real-time $T_{\rm obs}$, with acceleration search & $10$~min & $10$~min \\
    Max. real-time $T_{\rm obs}$, no acceleration search & $30$~min & $30$~min \\
    Time sampling & $100\;\upmu$s & $64\;\upmu$s \\
    \hline
    \end{tabular}
\end{table}


The configurable parameters are the gain employed and, correspondingly, the field of view that can be tiled out. These both depend on the choice of sub-array that is used for the search. The more elements we add into a sub-array, the better is the gain. However, as the tied-array beam sizes scale as the inverse square of the maximum baseline, there is a point where there are insufficient beams to cover the primary field of view. It is in this `sweet spot' that surveys are best performed.

Band 2 is an octave feed covering $950-1760$~MHz. For our Band 2 survey, we choose the conventional $1.4$~GHz as the centre frequency of a $300$-MHz bandwidth. When considering which parameters to choose for Band 1, which is a 3:1 band from $350-1050$~MHz, we examine the corresponding gain. As the gain improves with increasing frequency, we choose the highest frequency range within the band which allows for the most dishes. As the MeerKAT dishes will not have Band 1 feeds, but rather their UHF receivers covering $580-1015$~MHz~\citep{j16} we choose the top $300$~MHz of \textit{this} range, meaning a Band 1 central frequency of $865$~MHz.

\begin{figure}
   \centering
   \includegraphics[trim = {10mm 20mm 10mm 20mm}, clip, width=\hsize]{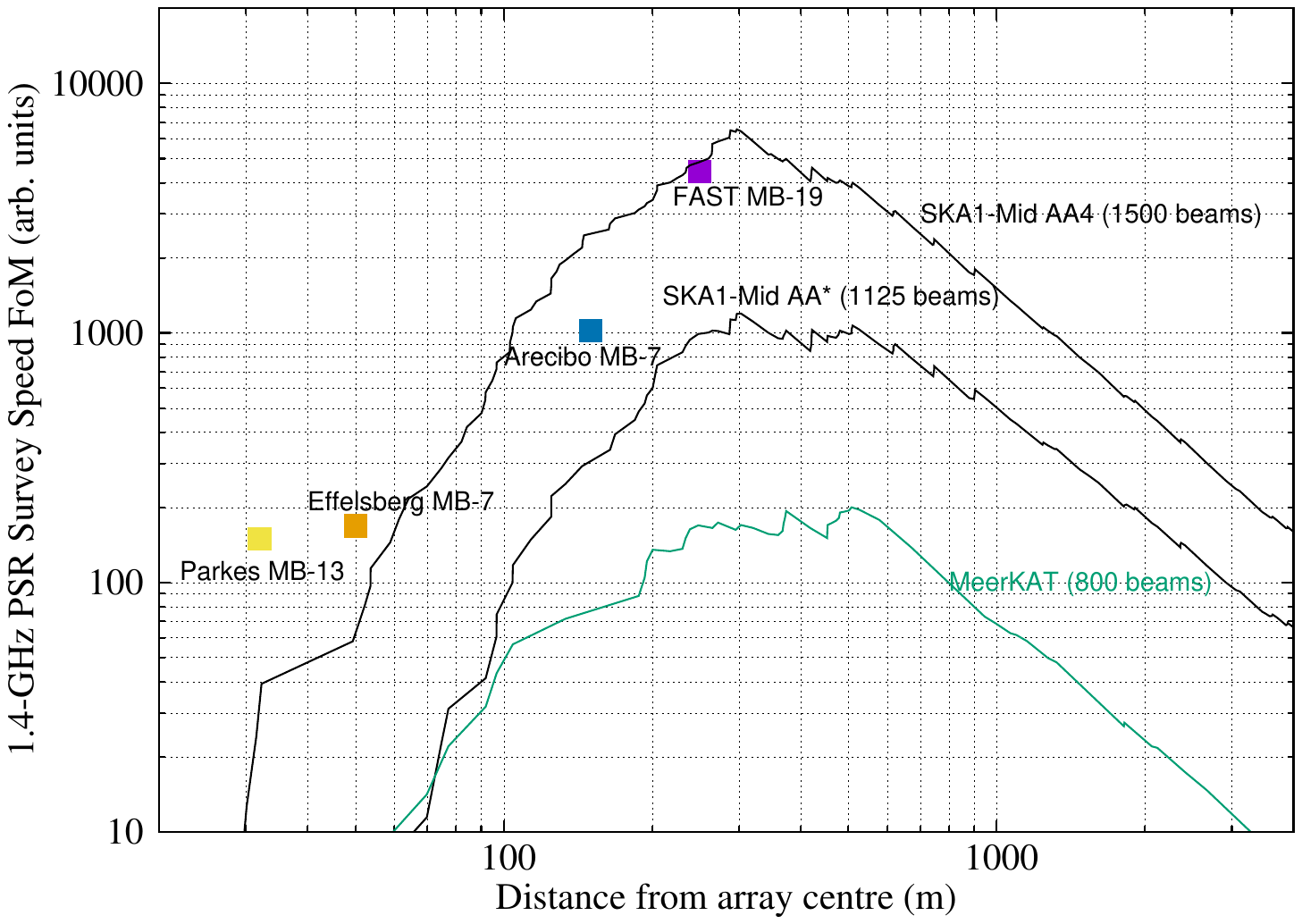}
   \caption{The $1.4$~GHz survey speed metric as a function of array radius for SKA1-Mid in the AA* and AA4 configurations, MeerKAT and several other radio telescopes for comparison. We use the resulting curves to model the optimal sub-array choice for untargeted pulsar searching for our pulsar yield analysis. For AA4, acceptable survey speed is found in the \textit{diameter} range $\sim 400\;\mathrm{m}$ to $\sim 1$~km; note that the horizontal axis shows \textit{radius}. Given the need to discover binary systems, we choose the inner $1$~km for our further analysis.}
   \label{fig:Mid_survey_speed}
\end{figure}

For AA4, the gain of the full SKA1-Mid array in Band 2 (Band 1) is $10.0~\mathrm{K/Jy}$ ($9.7~\mathrm{K/Jy}$) at $1.4$~GHz ($865$~MHz). For AA*, the corresponding full array gain in Band 2 (Band 1) is $7.0~\mathrm{K/Jy}$ ($6.8~\mathrm{K/Jy}$) at $1.4$~GHz ($865$~MHz). Smaller sub-arrays allow a much larger instantaneous field of view at the expense of gain. To find an optimal combination between these two parameters, we consider the $1.4$~GHz pulsar survey speed figure of merit, which is given by $\mathrm{min}(\mathrm{FoV}_{\rm primary},N_{\rm beams}\mathrm{FoV}_{\rm tied})A_{\rm eff}^2$. Figure~\ref{fig:Mid_survey_speed} shows the survey speed for SKA1-Mid as well as a few other instruments for reference. 

It is clear, for both AA* and AA4, that the $1.4$~GHz pulsar survey speed metric peaks at a radius of approximately $300$~m from the array centre, i.e., the inner $600$~m in diameter. Satisfactory survey speed is possible for sub-arrays ranging from the inner $\sim400$~m to the inner $1$~km, in diameter. We choose the inner $\sim1$~km in this work to allow for larger instantaneous gain which is needed both for discovering fainter objects but especially for finding accelerated binary systems. For the latter, the computational scaling of acceleration searches means that one cannot compensate for a lack in gain with more observing time, as computation costs scale with the cube of the observing time~\citep{kbk+15}.

The gain of SKA1-Mid was determined using a sensitivity calculator\footnote{\texttt{https://github.com/evanocathain/SKA}} used for previous analyses of the anticipated scientific performance of SKA1~\citep{k18,bbb+19}, updated for the current, final, array configurations. 
For AA4, the gain for the innermost $1$~km in Band 2 (Band 1) is $4.60\;\mathrm{K/Jy}$ ($4.40\;\mathrm{K/Jy}$) at $1.4$~GHz ($865$~MHz). The corresponding field of view that can be tiled out for such a sub-array is $0.62\deg^2$ ($1.62\deg^2$) in Band 2 (Band 1).

For AA*, the gain for the innermost $1$~km in Band 2 (Band 1) is $2.75\;\mathrm{K/Jy}$ ($2.60\;\mathrm{K/Jy}$) at $1.4$~GHz ($865$~MHz). The corresponding instantaneous field of view for AA* is unchanged from AA4 for a sub-array of the same size with the exception of the linear scaling of the number of tied array beams available. We ignore this factor however, essentially meaning we consider, for AA*, a survey with $(1500/1150)\sim 1.3\times$ more pointings than for AA4.

We note that the above gains correspond to effective areas for AA4 in Band 2 (Band 1) of $\sim0.013\;\mathrm{km}^2$ ($\sim0.012\;\mathrm{km}^2$) for this innermost $1$~km sub-array, equivalent to a fully-illuminated\footnote{A fully-illuminated dish is not realistic. Typically aperture efficiencies are $\eta\approx0.6$. This means that one might multiply the quoted equivalent dish diameters by $1/\sqrt{\eta}\sim1.3\times$ for a more realistic comparison.} dish of $\sim 127$~m ($\sim124$~m) diameter. For AA*, the corresponding values in Band 2 (Band 1) are an effective area of $\sim0.008\;\mathrm{km}^2$ ($\sim0.007\;\mathrm{km}^2$), equivalent to a fully-illuminated $\sim 98$~m ($\sim96$~m) dish.



\subsection{SKA1-Low Survey Parameters}
The PSS specifications for SKA1-Low AA4 (AA*) are that 500 (250) tied-array Stokes $I$ beams of $100$-MHz bandwidth are available 
for real-time Fourier domain acceleration searches and single pulse searches. As discussed above, the maximum pointing duration for this real-time performance is $10$~min if performing acceleration searches, $30$~min otherwise. As for Mid, we opt again for $10$~min pointings in our selection. Before we can perform the same exercise as above to choose the relevant search sub-array, we must consider the sub-band to use for a pulsar search from within the 7:1 band available. The PSS system can use $100$~MHz of band from \textit{anywhere} within the $50-350$~MHz range. In 2015, we had considered a $150-250$~MHz range but we now exclude the range $240-270$~MHz which is polluted with high-occupancy radio frequency interference (RFI)~\citep{swl16,swe17}, even at the Murchison Radio Observatory, one of the most radio quiet locations on Earth. Because of this, and even though RFI from satellites is also seen across the entire band at least a few percent of the time at all SKA1-Low frequencies~\citep{gts25}, we have chosen a frequency range of $140-240$~MHz for our SKA1-Low search range. Consequently, we examine the survey speed for SKA1-Low at a central frequency of $190$~MHz to determine its optimal sub-array size.

\begin{figure}
    \centering
    \includegraphics[trim={10mm 20mm 10mm 20mm},clip, width=\hsize]{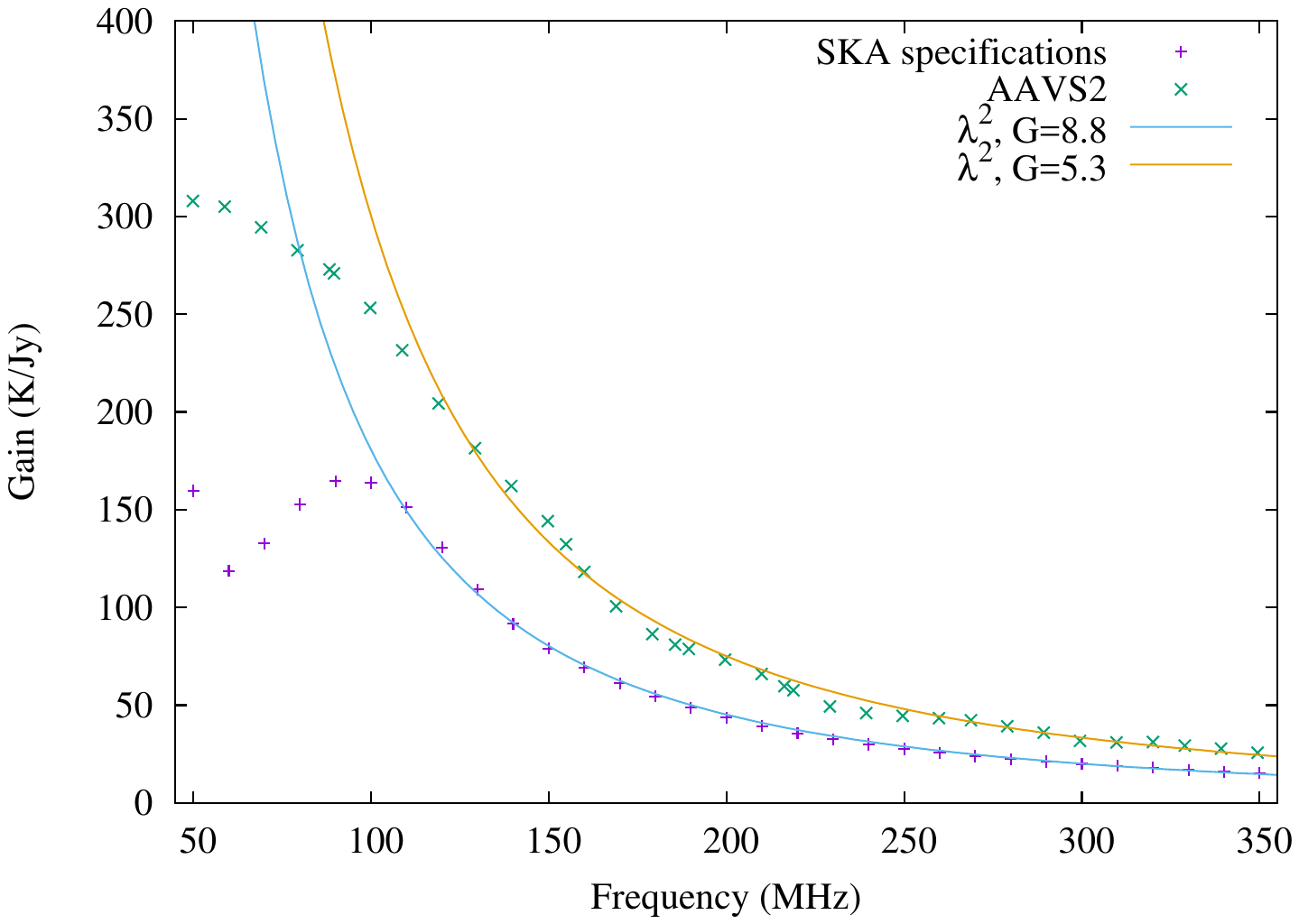}
    \caption{The gain at zenith of the full SKA1-Low AA4 array (purple plus signs) is shown, along with the corresponding SKA1 design specification (green crosses), which we can see is surpassed. This is based on the AAVS2 prototype; the SKALA4.1 response is likely even better. The conversion from $A_{\rm eff}/T_{\rm sys}$ to $A_{\rm eff}$ goes awry at the lowest frequency, $50$~MHz, as the system temperature balloons in a way not accounted for by the prescribed model, even though it allows for a factor of $2$ increase in $T_{\rm sky}$ from $60$~MHz to $50$~MHz. The effective area of an antenna is given as $\lambda^2G/(4\pi)$, where $G$ is the directivity or gain. For illustration, a fit to each is shown above the dense-sparse transition. It is clear from this that there is frequency dependence in the directivity.}
    \label{fig:low_sensitivity}
\end{figure}

In contrast to Mid, the sensitivity of Low is more complex because it: (a) is an aperture array; (b) has a large fractional bandwidth; and (c) observes to very low frequencies approaching the ionospheric cut off. The final on-sky performance of the SKALA4.1 antennas currently being rolled out is not known, but we take the performance of the Aperture Array Verification System 2 (AAVS2, \citealt{mpb+22}), the last in a series of prototype arrays, as our guide. For this work, we use a sensitivity calculator\footnote{\scriptsize{\texttt{https://github.com/marcinsokolowski/station$\_$beam}}} developed by \citet{std+22}. Figure~\ref{fig:low_sensitivity} shows the resulting effective area at zenith as a function of frequency for AAVS2, and compares this to the Level 1 system requirements for SKA1-Low~\citep{L1v12}. The latter are stated in $A_{\rm eff}/T_{\rm sys}$ but with a prescribed model for $T_{\rm sky}$~\citep{L1v12}, the largest component of $T_{\rm sys}$. We use this same sky temperature model, along with measured AAVS2 values for the temperature contributions from the low-noise amplifier and receiver components, i.e., $35$~K~\citep{b22} and $12$~K~\citep{b20}, respectively, to back out the SKA1 specification in terms of effective area. The gain has the expected frequency-dependent shape with effective area increasing $\propto \lambda^2$, i.e. increasing as frequency decreases until the frequency-dependent size of the individual antennas in a station reach the physical spacing. At lower frequencies than this so-called dense-sparse transition, the response is not flat as one might expect, for many reasons such as coupling interactions between antennas.

\begin{figure}
    \centering
    \includegraphics[trim={10mm 20mm 10mm 20mm},clip, width=\hsize]{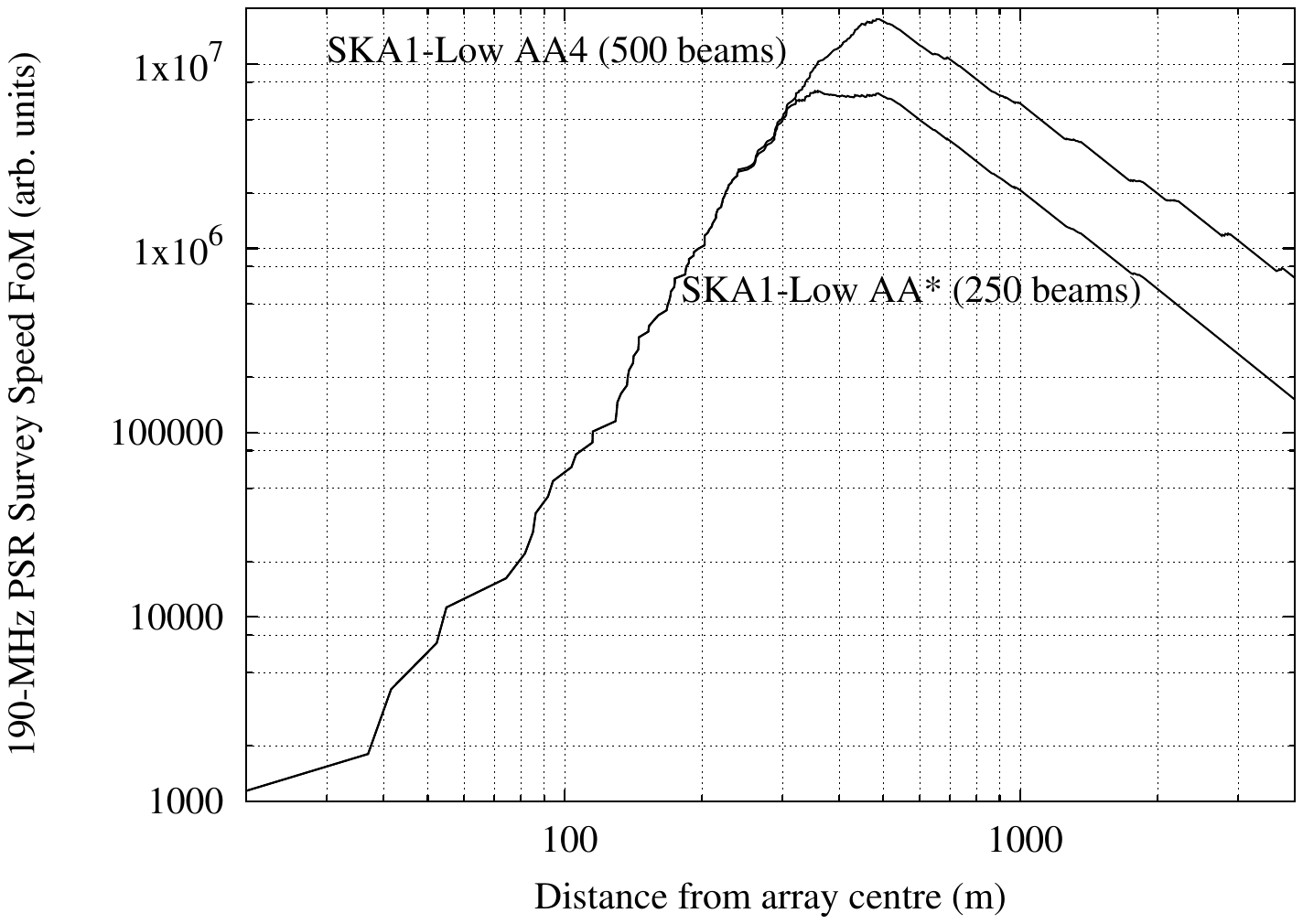}
    \caption{The survey speed of SKA1-Low as a function of radius from the centre of the array. Acceptable survey speed is found for the inner $\sim 600$~m to $1$~km diameter for AA*; for AA4 acceptable survey speed is available for inner $\sim 2$~km in diameter. It can be seen that, for pulsar search applications, the fact that the loss of stations between AA4 and AA* happens outside the core means the impact on the survey speed is far less dramatic than for SKA1-Mid.}
    \label{fig:Low_survey_speed}
\end{figure}

Figure~\ref{fig:Low_survey_speed} shows the survey speed for SKA1-Low at $190$~MHz for zenith. Here the differences between AA* and AA4 are less drastic than for Mid. Using the same logic as above for SKA1-Mid, we again choose a sub-array consisting of the inner $1$~km diameter for SKA1-Low. For AA4 it would also be possible to choose a sub-array consisting of the inner $\sim 2$~km, but this is not the case for AA* as the array thins out significantly after the inner kilometre, so for consistency of comparison we choose the inner $1$~km for our calculations in both cases. For AA4 (AA*), the inner $1$~km consists of $217$ ($192$) stations, for a gain of $33.4\;\mathrm{K/Jy}$ ($29.5\;\mathrm{K/Jy}$). We note that this corresponds to an effective area of $\sim0.092\;\mathrm{km}^2$ ($\sim0.082\;\mathrm{km}^2$) for this innermost $1$~km, equivalent to a fully-illuminated $\sim 340$~m ($\sim320$~m) dish. The gain off zenith is less than quoted above; we account for this in our estimates. As for Mid, one could phase up a much larger Low sub-array, for targeted observations --- the full array gain for the 512 (307) stations of AA4 (AA*) is $78.8\;\mathrm{K/Jy}$ ($47.2\;\mathrm{K/Jy}$).

\begin{figure*}[t!]
    \centering
    \begin{minipage}{.32\linewidth}
        \centering
        \includegraphics[width=\linewidth]{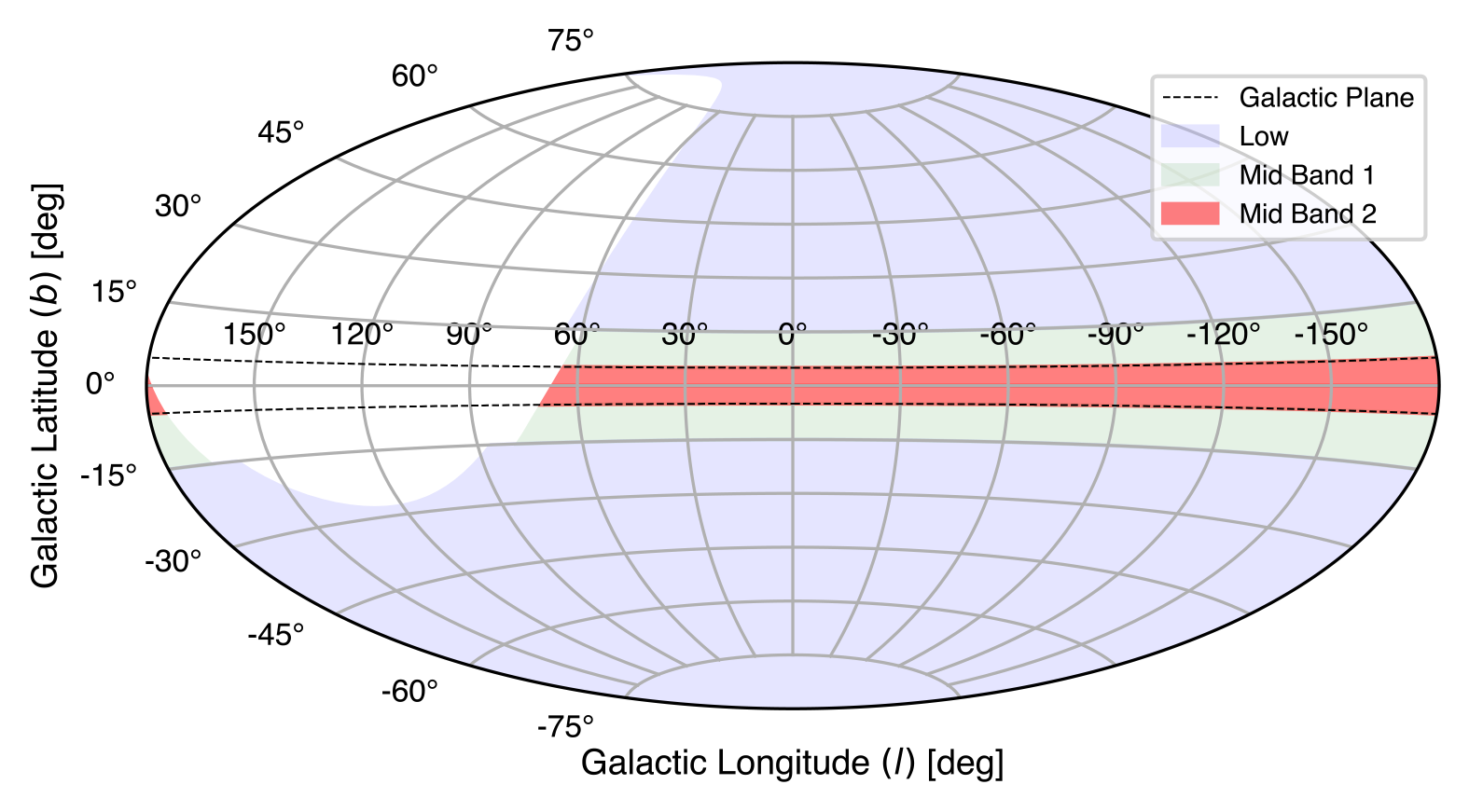}
    \end{minipage}
    \hfill
    \begin{minipage}{.32\linewidth}
        \centering
        \includegraphics[width=\linewidth]{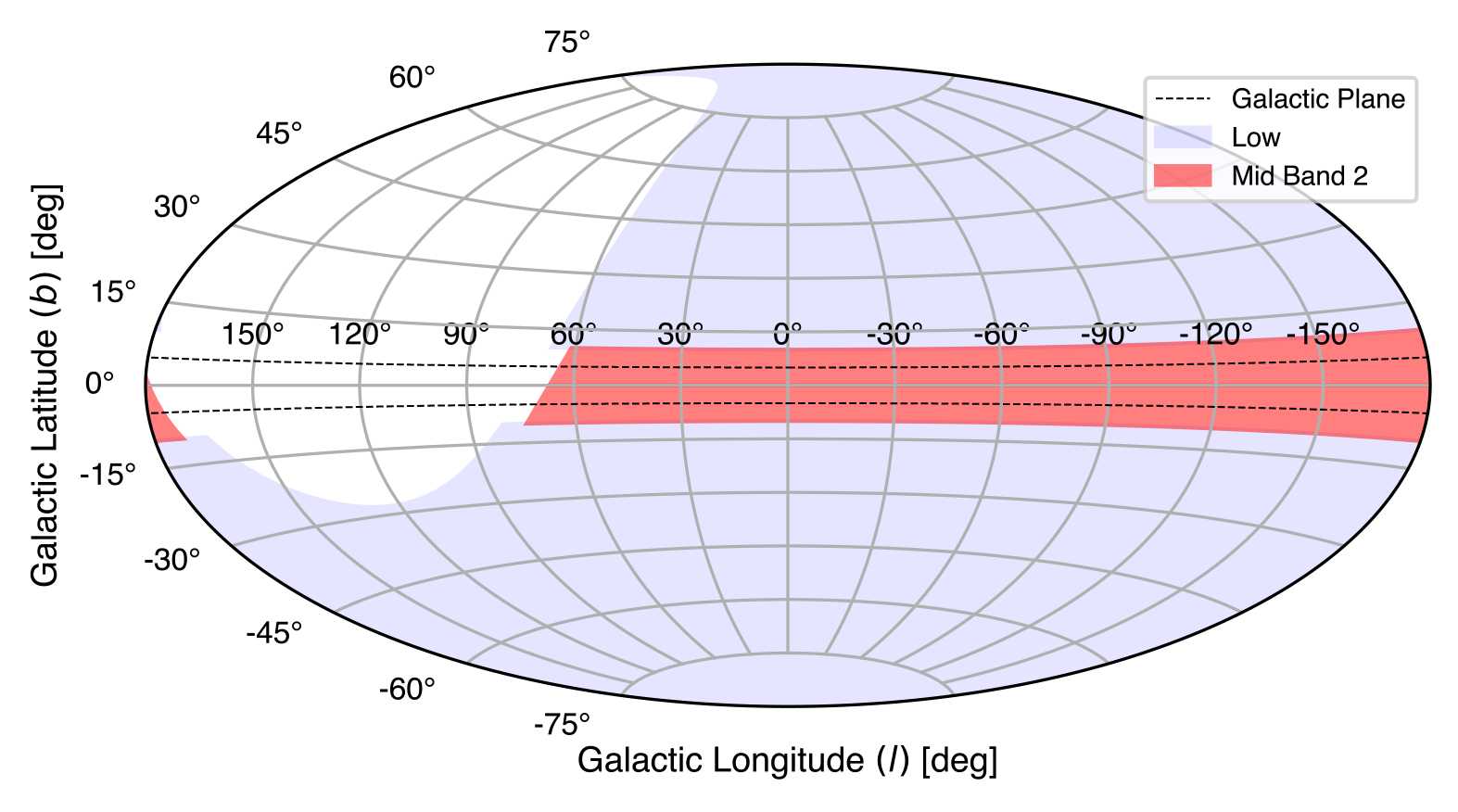}
    \end{minipage}
    \hfill
    \begin{minipage}{.32\linewidth}
        \centering
        \includegraphics[width=\linewidth]{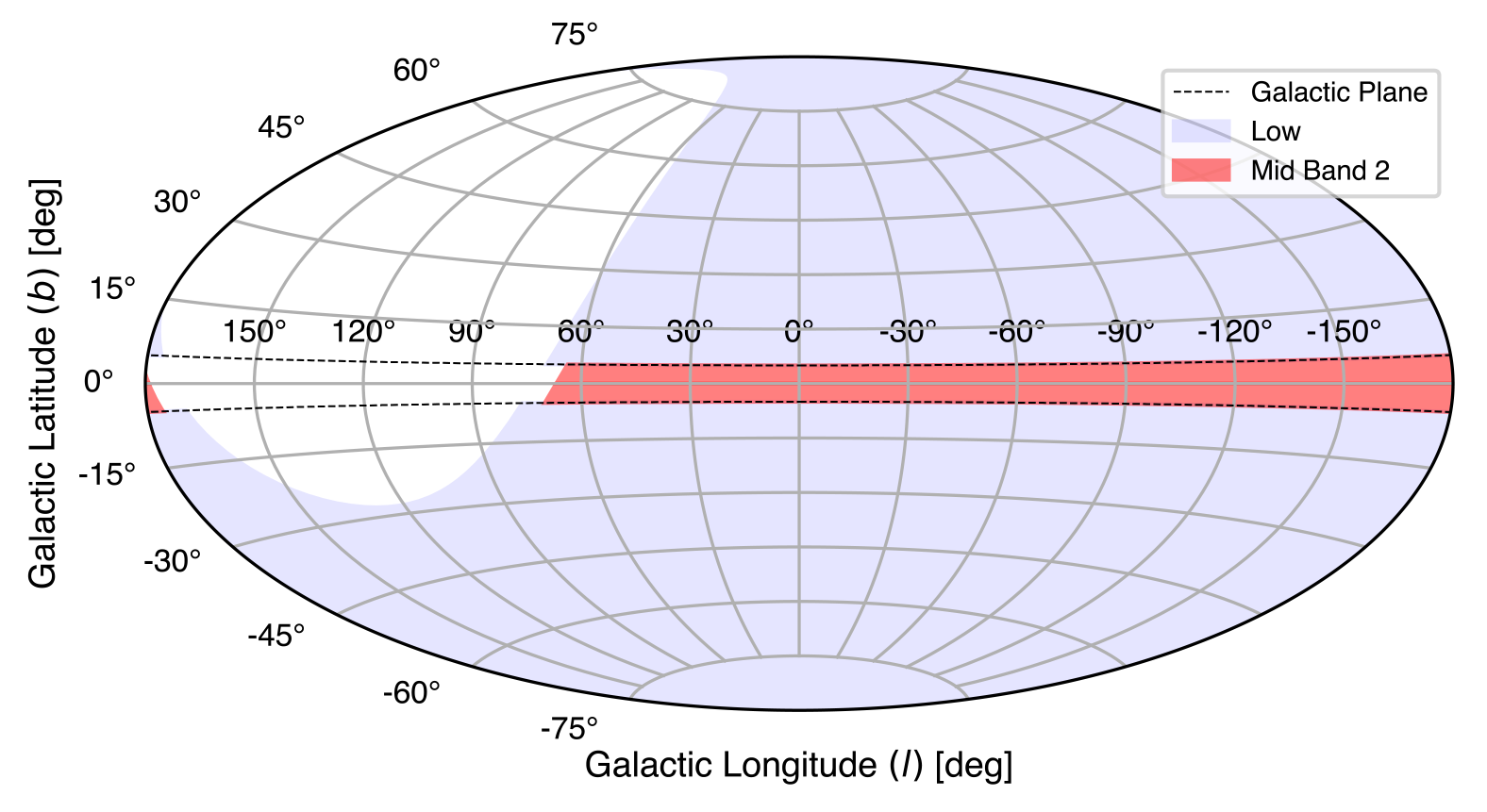}
    \end{minipage}
\caption{Aitoff projections in Galactic coordinates showing the sky coverage for the three illustrated composite survey options summarised in Table~\ref{tab:survey_options}. SKA1-Low coverage is marked in blue, SKA1-Mid Bands 1 and 2 are green and red respectively.}\label{fig:survey_options}
\end{figure*}

\begin{table*}
    \setlength{\tabcolsep}{1pt}
    \caption{The three survey options explored in this study with different balances between the three relevant bands for pulsar searches.}
    \label{tab:survey_options}
    \begin{minipage}{.31\linewidth}
        \centering
        \label{tab:first_table}
        \medskip
        \begin{tabular}{lc}
            \textbf{Survey Option 1} & \\
            Band & Latitude Range \\
            \hline
            Low & $|b|>15\deg$ \\
            Mid Band 1 & $15\deg > |b|>5\deg$ \\
            Mid Band 2 & $|b|<5\deg$ \\
            \hline
        \end{tabular}
    \end{minipage}
    \hfill
    \begin{minipage}{.31\linewidth}
        \centering
        \label{tab:second_table}
        \medskip
        \begin{tabular}{lc}
            \textbf{Survey Option 2} & \\
            Band & Latitude Range \\
            \hline
            Low & $|b|>10\deg$ \\
            Mid Band 1 & N/A \\
            Mid Band 2 & $|b|<10\deg$ \\
            \hline
        \end{tabular}    
    \end{minipage}
    \hfill
    \begin{minipage}{.31\linewidth}
        \centering
        \label{tab:third_table}
        \medskip
        \begin{tabular}{lc}
            \textbf{Survey Option 3} & \\
            Band & Latitude Range \\
            \hline
            Low & $|b|>5\deg$ \\
            Mid Band 1 & N/A \\
            Mid Band 2 & $|b|<5\deg$ \\
            \hline
        \end{tabular}    
    \end{minipage} 
\end{table*}

\section{Expected Yield of Pulsars}
In the following, we have considered pulsar surveys up to $+30\deg$ in declination for both SKA1-Mid bands, and up to $+36\deg$ in declination for SKA1-Low. We then implemented three different strategies for Galactic latitude coverage, as defined in Table~\ref{tab:survey_options} and as illustrated in Figure~\ref{fig:survey_options}. Each of these three composite surveys were unconstrained in Galactic longitude. In our analyses here the bands for each survey option do not overlap, so that each band detects unique pulsars. This means we do not have any duplicate sources detected in multiple bands. 


\subsection{Snapshot and evolutionary numbers}\label{sec:yields}
\begin{figure*}[t!]
    \centering
    \begin{minipage}{.32\linewidth}
        \centering
        \includegraphics[width=\linewidth]{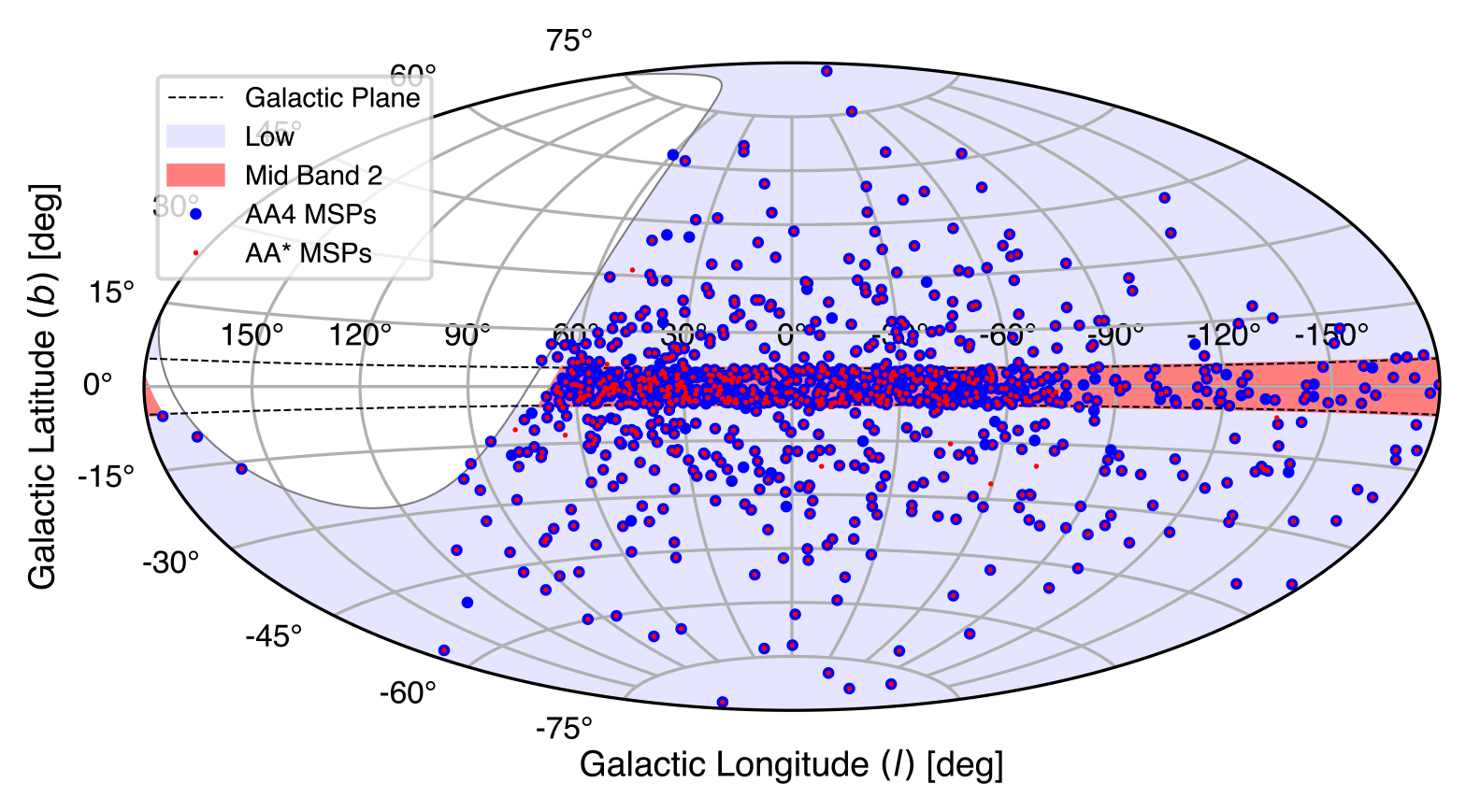}
    \end{minipage}
    \hfill
    \begin{minipage}{.32\linewidth}
        \centering
        \includegraphics[width=\linewidth]{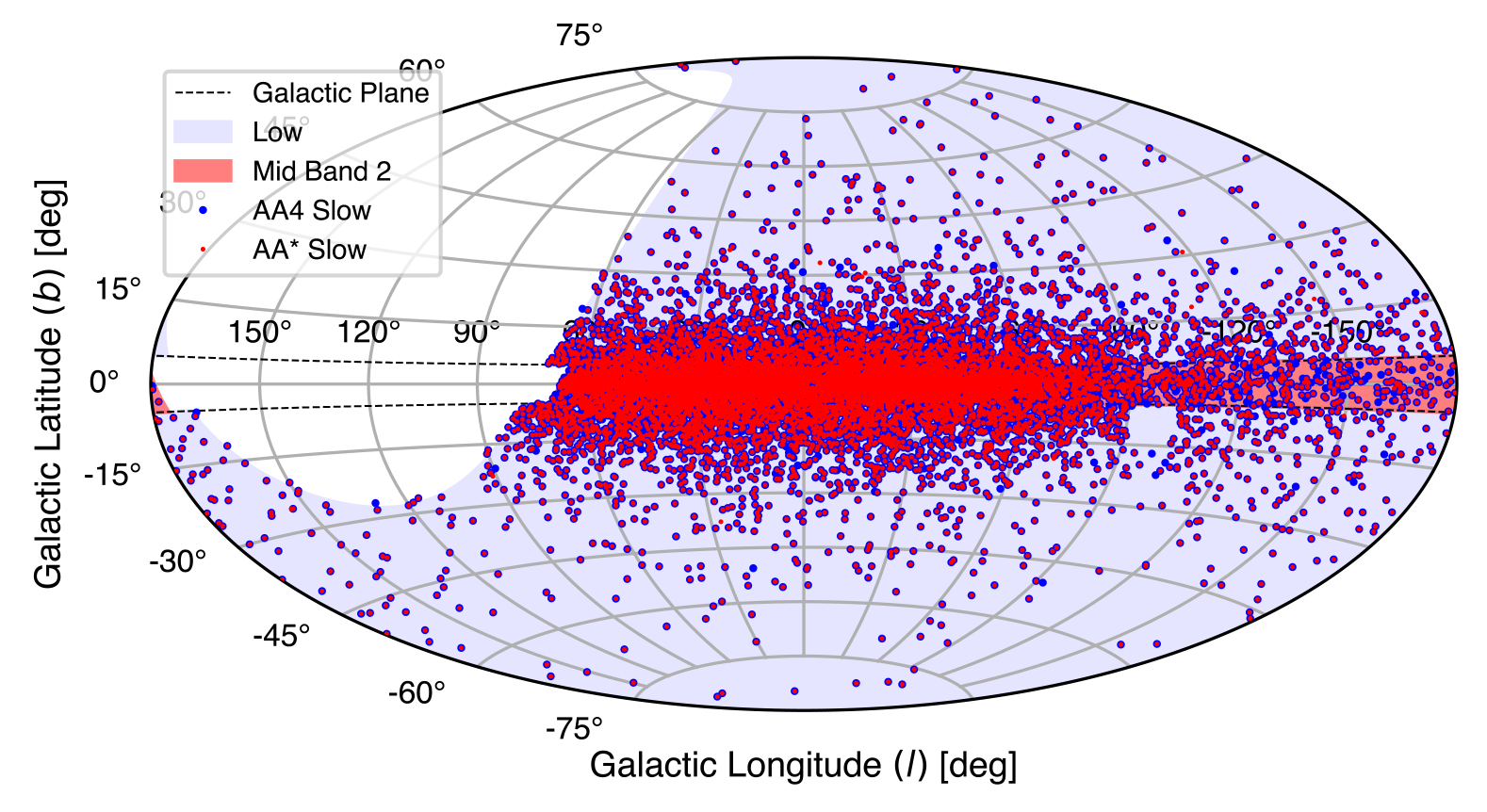}
    \end{minipage}
    \hfill
    \begin{minipage}{.32\linewidth}
        \centering
        \includegraphics[width=\linewidth]{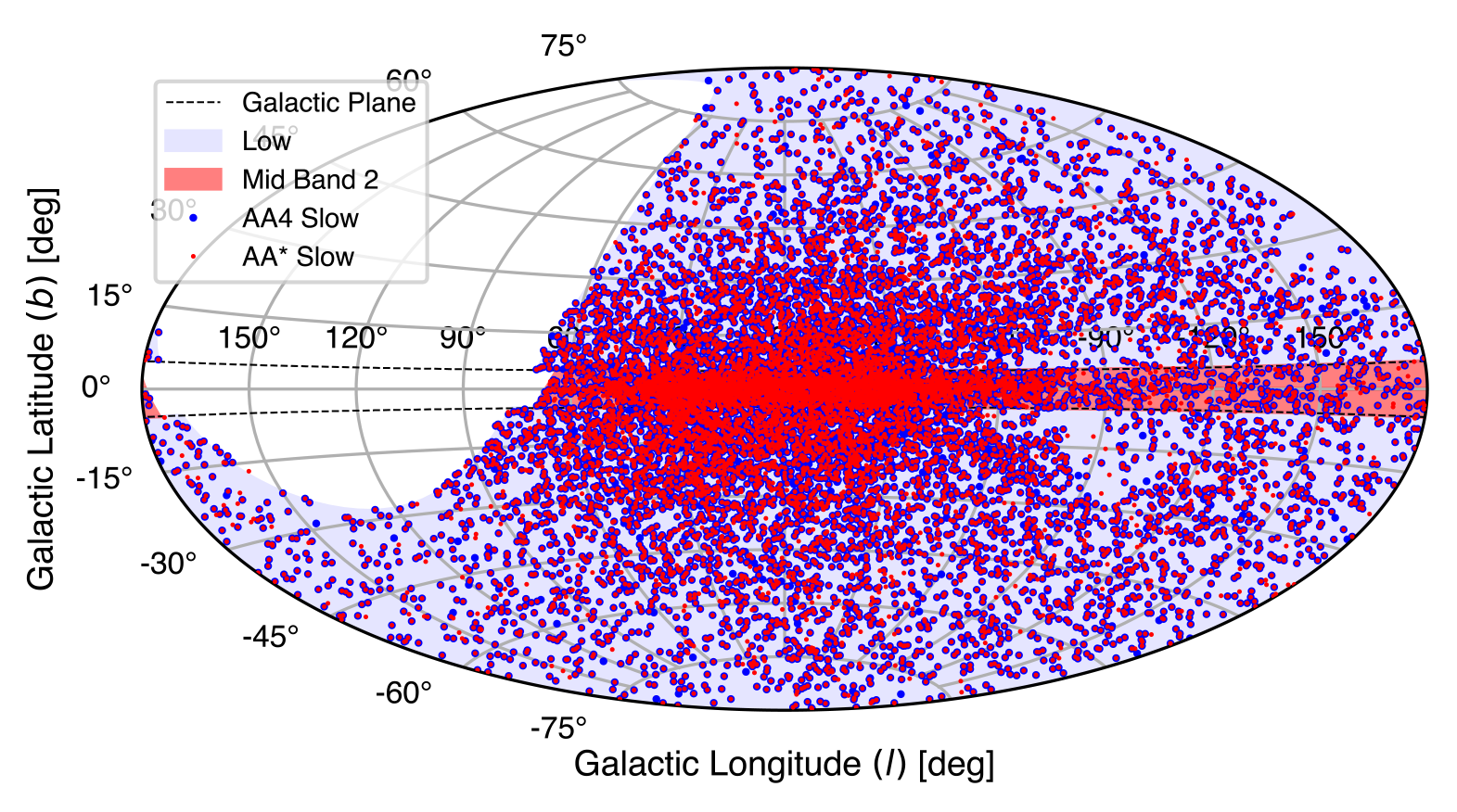}
    \end{minipage}
\caption{Aitoff projections in Galactic coordinates for Survey Option 3 showing the pulsar yield for a randomly chosen, but representative, realisation from our analyses. The left panel shows the MSPs predicted by the snapshot method; the middle panel shows the slow pulsars for the snapshot method; the right panel shows the slow pulsars from the evolutionary method. In each case, AA* and AA4 yields are highlighted. We note that the pulsars detected in AA* are a subset of those detected in AA4, i.e., AA* detections are shown in red, AA4 detections are both red and blue. The two slow pulsar distributions show the clear distinction in the scale height between the snapshot and evolutionary approaches, as discussed in the main text.}\label{fig:survey_yields}
\end{figure*}

We have codified the above telescope and SKA1 survey parameters to define surveys for analysis with \textsc{psrpoppy} and the evolutionary approach outlined in \S~\ref{sec:pop_sims}. The results using the snapshot methodology, leading to counts for slow pulsars as well as MSPs, are shown in Table~\ref{tab:psrpoppy_yields} for our three survey options. Table~\ref{tab:evol_yields} summarises the corresponding yields of the evolutionary approach. The latter are for slow sources only as the framework is unsuited to modelling fast MSPs. Both tables show the means and standard deviations of the yields, with Table~\ref{tab:psrpoppy_yields} being based on $100$ iterations, whereas Table~\ref{tab:evol_yields} quotes values based on $10$ simulation runs. Standard deviations are notably higher in the snapshot approach because each simulation run samples from all the underlying distributions that enter the model. In the evolutionary approach, this is not possible due to the computational cost associated with evolving pulsar parameters over time. For feasibility, we have evolved a single underlying population and then applied the detection pipeline $10$ times. As a result, stochasticity in the evolutionary results arises only from variability in survey modelling rather than in the underlying population itself. To illustrate the projected distribution of discoveries for these estimates, Figure~\ref{fig:survey_yields} shows an example realisation for both approaches. 

\begin{table*}[!t]
    \setlength{\tabcolsep}{1pt}
    \caption{Shown are the expected yields for the maximum number of pulsars one could detect using the AA* and AA4 configurations for each of three composite survey strategies as determined with our snapshot approach. We have not imposed any overall observing time constraints at either telescope. Numbers quoted are the rounded means from $100$ iterations and numbers in parentheses represent the standard deviation of the values from those $100$ iterations. As presented here each band covers a unique patch of sky so that one simply adds the yield from each band to determine the total. Covering the same patch of sky with more than one band increases the yield over what is stated here.}
    \label{tab:psrpoppy_yields}
    \begin{minipage}{.31\linewidth}
        \centering
%
%
%
        \begin{tabular}{lcc}
            \textbf{Survey Option 1} & \\
            Band & Slow PSRs & MSPs \\
            \hline
            \textbf{AA4} & & \\
            Low & 1620(70) & 280(20) \\
            Mid Band 1 & 3300(100) & 440(20) \\
            Mid Band 2 & 8150(240) & 560(30) \\
            TOTAL & 13070 & 1280 \\
            \hline
            \textbf{AA*} & & \\
            Low & 1570(60) & 270(20) \\
            Mid Band 1 & 2280(80) & 310(20) \\
            Mid Band 2 & 5400(150) & 380(20) \\
            TOTAL & 9250 & 960 \\
            \hline
        \end{tabular}
    \end{minipage}
    \hfill
    \begin{minipage}{.31\linewidth}
        \centering
%
%
%
        \begin{tabular}{lcc}
            \textbf{Survey Option 2} & \\
            Band & Slow PSRs & MSPs \\
            \hline
            \textbf{AA4} & & \\
            Low & 2900(100) & 370(20) \\
            Mid Band 1 & - & - \\
            Mid Band 2 & 10800(300) & 950(30) \\
            TOTAL & 13700 & 1320 \\
            \hline
            \textbf{AA*} & & \\
            Low & 2800(100) & 350(20) \\
            Mid Band 1 & - & - \\
            Mid Band 2 & 7300(200) & 640(20) \\
            TOTAL & 10100 & 990 \\
            \hline
        \end{tabular}    
    \end{minipage}
    \hfill
    \begin{minipage}{.31\linewidth}
        \centering
%
        \begin{tabular}{lcc}
            \textbf{Survey Option 3} & \\
            Band & Slow PSRs & MSPs \\
            \hline
            \textbf{AA4} & & \\
            Low & 5250(170) & 470(20) \\
            Mid Band 1 & - & - \\
            Mid Band 2 & 8150(240) & 560(30) \\
            TOTAL & 13400 & 1030 \\
            \hline
            \textbf{AA*} & & \\
            Low & 4990(170) & 440(20) \\
            Mid Band 1 & - & - \\
            Mid Band 2 & 5400(150) & 380(20) \\
            TOTAL & 10390 & 820 \\
            \hline
        \end{tabular}    
    \end{minipage}
\end{table*}

\begin{table*}[!t]
    \setlength{\tabcolsep}{1pt}
    \caption{Expected yields for the maximum number of pulsars we could detect in the AA* and AA4 configurations for our three survey strategies as obtained with the evolutionary simulations. Again no observing time constraint is imposed. Numbers quoted are the rounded means from $10$ iterations based on a fixed evolved population. Numbers in parentheses represent the standard deviation of the values from those $10$ iterations. Stochasticity arises from uncertainty in the detection implementation only and not the underlying population model. Note that the evolutionary framework outlined above does not allow us to model MSPs. The first set of numbers for each survey option is a bare count of all detected sources, while the latter only counts those sources that lie above a pulsar death line (DL) based on an extreme twisted, multi-polar magnetospheric configuration.}
    \label{tab:evol_yields}
    \begin{minipage}{.31\linewidth}
        \centering
        \begin{tabular}{lcc}
            \textbf{Survey Option 1} & \\
            Band & full count & above DL \\
            \hline
            \textbf{AA4} & & \\
            Low & 6320(20) & 5420(20) \\
            Mid Band 1 & 1920(10) & 1850(10) \\
            Mid Band 2 & 3420(20) & 3360(10) \\
            TOTAL & 11660 & 10630 \\
            \hline
            \textbf{AA*} & & \\
            Low & 5750(20) & 4980(20) \\
            Mid Band 1 & 1350(10) & 1310(10) \\
            Mid Band 2 & 2570(10) & 2540(10) \\
            TOTAL & 9670 & 8830 \\
            \hline
        \end{tabular}
    \end{minipage}
    \hfill
    \begin{minipage}{.31\linewidth}
        \centering
        \begin{tabular}{lcc}
            \textbf{Survey Option 2} & \\
            Band & full count & above DL \\
            \hline
            \textbf{AA4} & & \\
            Low & 7580(30) & 6570(20) \\
            Mid Band 1 & - & - \\
            Mid Band 2 & 4730(20) & 4630(20) \\
            TOTAL & 12310 & 11200 \\
            \hline
            \textbf{AA*} & & \\
            Low & 6900(20) & 6040(10) \\
            Mid Band 1 & - & - \\
            Mid Band 2 & 3490(10) & 3440(10) \\
            TOTAL & 10390 & 9480 \\
            \hline
        \end{tabular}    
    \end{minipage}
    \hfill
    \begin{minipage}{.31\linewidth}
        \centering
        \begin{tabular}{lcc}
            \textbf{Survey Option 3} & \\
            Band & full count & above DL \\
            \hline
            \textbf{AA4} & & \\
            Low & 8830(20) & 7750(20) \\
            Mid Band 1 & - & - \\
            Mid Band 2 & 3420(20) & 3360(20) \\
            TOTAL & 12250 & 11110 \\
            \hline
            \textbf{AA*} & & \\
            Low & 8050(30) & 7110(20) \\
            Mid Band 1 & - & - \\
            Mid Band 2 & 2570(10) & 2540(10) \\
            TOTAL & 10620 & 9650 \\
            \hline
        \end{tabular}    
    \end{minipage}
\end{table*}

We note the very important caveat that in each case, the resulting counts are the \textit{maximum} number of pulsars we might \textit{detect} with AA* and AA4 in the three survey options, so as to state the full potential. To estimate \textit{discoveries}, in the case of slow pulsars we would simply subtract the known pulsars in the sky region. For MSPs, this would result in an over-subtraction as most MSPs have not been found in blind surveys but rather in deep targeted searches, e.g., of \textit{Fermi} point sources~\citep{rrc+11,cck+16}. Indeed SKA will also perform long targeted searches~\citep{Abbate2025_SKA_GalCen,Bagchi2025_SKA_GlobClust} so that the overall MSP numbers will be higher. We note further that the timing precision afforded by SKA in observing these MSPs will be considerably better: pulsar timing array sensitivity scales with both the number of MSPs but also the timing precision~\citep{Shannon2025_SKA_SKAPTA}.

In all survey options, we find that the AA* configuration yields systematically fewer detected sources than AA4. The differences are less pronounced in the Low band (expected based on the discussion around Figure~\ref{fig:Low_survey_speed}) with AA* typically achieving over $90\%$ of the AA4 yield, but become more significant for SKA1-Mid, where the AA* yield drops to $\sim 70\%$ of the corresponding AA4 value in both bands. While the evolutionary approach does not provide information on MSPs, it allows us to explicitly track the period and period derivative evolution of each pulsar as it ages---an example $P$-$\dot{P}$ diagram for Survey Option 3 is shown in Figure~3 in \citet{Levin2025_SKA_NSpop}. Importantly, the evolutionary approach outlined in \S \ref{sec:pop_evl} does not impose an artificial death line below which the pulsar radio emission switches off, as this was not required to reproduce the calibration surveys outlined above (see \citealt{grpn24} for details). However, a death line may become relevant for SKA1-era surveys based on the telescopes' increased sensitivity and SKA1-Low's wide sky coverage. To estimate the death line's potential impact, we applied a death line criterion based on a twisted multipolar magnetospheric model \citep{cr93,rhp+24}. This additional filter further reduces the Low counts, where older, lower-$\dot{P}$ pulsars dominate, while having a smaller effect on the Mid bands. The `above DL' columns in Table~\ref{tab:evol_yields} provide the yields that are most directly comparable to those of the snapshot simulations, which were calibrated against existing surveys at low latitudes and are therefore unlikely to predict many sources below the death line.

Comparing the total counts for each survey option and array assembly, we find that these totals are broadly consistent between the snapshot and evolutionary approaches. However, an important difference emerges in the relative contributions of the SKA1-Low and SKA1-Mid bands. In the snapshot approach, the Mid bands (particularly Band 2) consistently dominate the total yield, contributing a larger fraction of detected sources than Low across all options. In contrast, the evolutionary simulations predict the opposite trend: Low detections always exceed those in Mid Band 1 and Band 2. Quantitatively, in the evolutionary approach, Low detections account for $\sim 50-70\%$ of the total yield depending on the option, while in the snapshot simulations they contribute only $\sim15-40\%$. This discrepancy arises despite both approaches using the same underlying spectral index distribution and identical survey parameters, and reflects differences in the assumed physics of the underlying populations, such as the luminosity prescription and the spatial distribution of evolved pulsars. We explore these differences in greater detail below.


\subsection{Survey considerations}\label{sec:considerations}

When presenting our yields, we have not attempted to optimise for the relative observing time cost of each survey, and the related Galactic latitude coverage of each component of the composite search. The relative cost of a survey of a given area of the sky is $2.7$ times higher for SKA1-Mid Band 2 versus SKA1-Mid Band 1. The relative cost between SKA1-Mid Band 2 and SKA1-Low is a dramatic $55$ times, for AA4. For AA* the relative costs are slightly less stark as in that case Low has $50\%$ less tied-array beams, whereas Mid loses only $25\%$ of its tied-array beams relative to AA4. This weighting is only partially accounted for in our Galactic latitude constraints. One impact of this is that the maximum yield for SKA1-Mid Band 2 is unlikely to be realised given how competitive telescope access will be, but we wished to present the maximum potential numbers here. Conversely, a truly all-sky search with SKA1-Low is quite likely, even covering the Galactic plane, even where scattering and dispersion smearing will reduce the horizon of any survey. The Low survey speed is so much faster that the relative cost of also covering the plane is very low in comparison to Mid. Another way to state this is that Survey Option 3 is more likely closer to what will happen than Survey Option 2, which is, in turn, more likely than Survey Option 1. In a similar way, with the need for additional funding to realise AA4, the AA* numbers are also more likely to be closer to reality. Thus, Survey Option 3 for AA* with $\sim 10,000$ slow pulsars and $\sim 800$ MSPs to be detected---split roughly equally between Low and Mid Band 2 in the snapshot simulations, while the former count is roughly 2.5 times that of the latter in the evolutionary scenario for slow pulsars---is the most likely scenario from those presented here. When the time comes to perform the surveys with AA*/AA4, any and all such real-world constraints on availability can readily be incorporated. In the first instance, smaller-scale pilot surveys are likely to ensure the mechanics of observing, processing and analyses are fully tuned. This will allow verification and calibration of our estimates ahead of a large scale up. With the subset of the full arrays relevant to pulsar surveys being the innermost $1$~km, this work may be possible to commence earlier than for other science uses of the SKA. 

Another parameter which is not considered in our analyses but which will have significant real-world impact is the affect of RFI~\citep{mjs22}. The most optimistic case is where we either have time ranges and/or frequency ranges which are unusable so that we effectively have less observing time and/or bandwidth. This kind of RFI can be well mitigated with masking~\citep{mlc+01,kbr10}. Unfortunately, the spectral situation has changed in the past decade such that it is no longer the case that we have some frequencies that are unusable all the time. Instead, we now have a situation where \textit{all frequencies are bad some of the time}. This means that very significant effort needs to be put into RFI flagging in streaming data. Much work has been performed in this domain on MeerKAT~\citep{rsw+20,mjs22} and will be applied in the PSS systems on SKA1-Mid and SKA1-Low. One lesson learned from MeerKAT is that the UHF band (which overlaps with SKA Band 1) is cleaner than the L band (which overlaps with SKA Band 2), meaning that we should expect the SKA1-Mid Band 2 to be more affected by RFI than SKA1-Mid Band 1. This issue too, once we have some initial on-sky data with SKA1-Mid, will weigh into our final survey optimisation strategies. A counter point to this is that we might want to ensure we find as many Band 2 pulsars as possible, even given the larger relative cost in observing time. The MSPs so found would be the most valuable for pulsar timing array applications, as it is in Band 2, and in the future Band 3 which will operate in S band, where the highest timing precision is possible~\citep{Shannon2025_SKA_SKAPTA}. There are additionally many other possible considerations which we might employ, e.g., prioritising some of the FAST sky~\citep{hww+21} for cross-calibration purposes.

We reiterate that we present the \textit{maximum} possible number of
pulsar detections one could make for our chosen illustrative
configurations and surveys. We have not specified the total observing
time for these as there are still many parameters one could tweak
before performing these surveys which could dramatically effect these
numbers. Some choices of configuration are informed by observations,
e.g. the real on-sky performance and RFI occupancy seen in pilot
surveys; others depend on the final SKA build. For instance,
considering what LOFAR and MeerKAT have been capable of for many years
already, we are optimistic that SKA1-Mid (or SKA1-Low) will exceed
$1500 \times 300$-MHz ($500 \times 100$-MHz) tied-array beams for
pulsar search. With more beams the yields remain as above, but the
survey speed to find these pulsars decreases. At the same time, we
might expect to maintain the same survey speed in case more beams
become available and instead increase the size of the sub-array. This
would consequently lead to a larger instantaneous sensitivity,
resulting in an increased MSP yield (while the slow pulsar counts
would remain the same). Here, we do not find it useful to go beyond the illustrative surveys
outlined above and discuss the many additional possibilities that
could affect corresponding estimates, as these parameter choices will
not have crystalised before SKA1 commences full operations.

\section{Discussion \& Conclusion}\label{sec:discussion}
One can use the yield projections above to estimate the number of sub-populations that the SKA will detect. We take double neutron stars binaries as an example~\citep{tkf+17}, as this number is an important ingredient in investigation of the neutron star equation of state using SKA~\citep{Basu2025_SKA_EOS} and tests of gravity~\citep{VenkatramanKrishnan2025_SKA_Gravity}. Scaling from the known pulsar population using the period distribution of double neutron stars and of detections from Survey Option 3, one estimates $\sim 140$ ($\sim 110$) double neutron star systems to be detected with AA4 (AA*). 
%
%
%

In trying to estimate the uncertainty of such estimates we note that the total yield numbers presented in \S~\ref{sec:yields} have associated uncertainties that are dominated by systematics. Both methods endeavour to recreate the same underlying population of neutron stars but clearly make differing predictions. While one might look at the total predicted yields to estimate the systematic uncertainties, e.g., compare 13070 to 10630 for AA4 Survey Option 1, or compare 10390 to 9650 for AA* Survey Option 3, as described in \S~\ref{sec:considerations}, the deviations are larger when focusing on just Low or just one of the Mid bands. While some degree of uncertainty is inevitable as we are modelling a complex population, extrapolating beyond currently known population sizes, and estimating the performance of an as-yet-incomplete system, it is possible to rein in some of these uncertainties, even with the information at hand, well before SKA is fully operational. 

One key step toward improving yield predictions for pulsar detections with the SKA (and any other instrument) would be more accurate and complete reporting of results from existing pulsar surveys. In particular, information on blindly detected pulsars, i.e., those discovered without prior timing knowledge, is often incomplete. The distinctions between being blindly detected in a Fourier-Transform based search, versus being detected in a phase-coherent folding approach, discovery signal-to-noise ratios versus tuned values etc. are all crucial in properly modelling pulsar survey outputs. This information is often, but not always, available in original survey papers, and it is typically missing from the ATNF Pulsar Catalogue\footnote{\texttt{https://www.atnf.csiro.au/research/pulsar/psrcat/}}~\citep{mhth05}, which remains the primary input source for pulsar population synthesis. This limitation is one of our main reasons for selecting the various Parkes surveys discussed in \S~\ref{sec:pop_sims} (albeit different between the snapshot and evolutionary approaches), as they represent some of the most thoroughly searched and transparently reported surveys to date. In doing so, we are eschewing information from more recent surveys but with the completeness of the searches, and reporting thereof being unclear, we choose this conservative approach. For the evolutionary modelling framework in particular, having access to more than just the detection numbers is critical. Detailed reporting of each pulsar’s period and period derivative along with flux measurements is especially important when inferring the pulsars’ luminosity distribution~\citep{cbo20,prgr25}. This distribution is a key element in population synthesis, and different implementations have been employed in the literature \citep[e.g.,][]{fk06,cbo20, pkj+23, grpn24, sn24, prgr25}. Radio flux densities are generally inconsistently reported, causing difficulties in distinguishing between approximate estimates and well-calibrated measurements. For instance, in their rigorous flux density measurements of over 400 pulsars, \citet{jsk+15} identify several tens of pulsars with as high as $5\sigma$ deviations between their reported and measured values. Similarly many LOFAR flux densities determined from tied-array measurements, are seen to be uncertain by a factor of $2$ or more when compared with imaging observations~\citep{kvh+16, mkgm24}. Uniform and systematic reporting of radio flux densities together with rotational properties would allow for more reliable comparisons across the full pulsar population, helping to constrain luminosity-related parameters more robustly. We will also require this information to fully understand the population of pulsars found by SKA --- these sources will require timing \textit{by the SKA} to be fully characterised~\citep{Levin2025_SKA_NSpop}, not simply found and discarded, nor timed by other facilities. 

In the snapshot approach outlined in \S~\ref{sec:snapshot}, we chose the Parkes Southern Pulsar Survey ($420-452$~MHz), the Parkes Multibeam Pulsar Survey ($1230-1518$~MHz), and the Methanol Multibeam Survey ($6303-6879$~MHz) in part to span a broad range of observing frequencies. This matters because pulsar detection rates are sensitive to the underlying spectral index distribution. The spectral model we adopt here (see Figure~\ref{fig:spectra}), of a power-law spectrum for every pulsar, is likely an over-simplification, especially at SKA1-Low frequencies~\citep{jsk+15}. However, for our analyses we are able to find a solution with such a model that is consistent with the yields of all the surveys considered in this work. Incorporating additional surveys at other frequencies, such as the Green Bank North Celestial Cap (GBNCC) survey at $350\,$MHz \citep{slr+14,kmk+18}, would allow for further constraints on the spectral properties of the population. While it is possible that no single spectral index law describes the full population, analysis across a broad frequency range can help characterise sub-populations and refine model assumptions. SKA’s broad frequency coverage across SKA1-Mid and SKA1-Low will ultimately be key in determining the neutron stars' intrinsic spectral index distribution(s), and in particular similarities and differences across the slow pulsar and millisecond pulsar populations \citep{kjp+24}.

In contrast, the three surveys used in the evolutionary approach (namely the Parkes Multibeam Pulsar Survey, the Swinburne Intermediate-latitude Pulsar Survey, and the low- and mid-latitude High Time Resolution Universe survey, all at $1.4\,$GHz) were selected for their coverage across a wide range of Galactic latitudes, rather than frequencies. This is particularly important when modelling the neutron star's evolution from birth to present, as pulsars born in the Galactic plane gradually migrate to higher latitudes due to natal kicks. Surveys at higher latitudes, thus, probe older pulsars, which have evolved to lower period derivatives. These sources are important for understanding late-stage pulsar evolution, including magnetic field decay, beaming behaviour, and the potential ceasing of radio emission near the `death line'~\citep[e.g.,][]{cr93}. For the $P$-$\dot{P}$ diagrams corresponding to Survey Option 3, see Figure 3 in \citealt{Levin2025_SKA_NSpop}. SKA1-Low will be especially useful for detecting these evolved pulsars. In a sense, the pulsar discoveries made by SKA1-Low will enable the death line drop-off to be `seen'. Moreover, coverage across different latitudes informs models of both pulsar birth locations and supernova kick velocities. While the $z$-distribution of neutron star progenitors is relatively well constrained, the kick distribution remains less certain. Our evolutionary simulations are based on the often-used Maxwellian distribution from \citet{hllk05}. However, this analysis is now not only outdated as improved proper motion measurements have become available in the past two decades \citep[e.g.,][]{gsf+11,rgg+21,dds+23,sbf+24}, but the single Maxwellian distribution from \citeauthor{hllk05} may also incorrectly represent the pulsar's natal kick velocities, potentially overestimating them \citep[e.g.,][]{vic17,i20,mm20,mi23,dm25}.

The treatment of the kick velocity, which ultimately controls the vertical distribution of observed pulsars in the Galaxy, has a direct impact on our predicted SKA yields. Notably, the difference in SKA1-Low versus SKA1-Mid counts between the snapshot and evolutionary frameworks presented in Tables~\ref{tab:psrpoppy_yields} and \ref{tab:evol_yields}, respectively, may stem from differences in the spatial modelling. The \textsc{psrpoppy}-based snapshot model uses an exponential $z$-distribution with a $330\,$pc scale height for the slow pulsar population, which was calibrated using low-latitude surveys \citep{lfl+06}. This choice may bias the inferred number of pulsars at both low and high latitudes, potentially overestimating the pulsars at low latitudes (detected by Mid) and overestimates those at larger latitudes (detected by Low) relative to the evolutionary simulations. Similar issues might affect the distribution of millisecond pulsars for which we cannot compare to an evolutionary model. The latter is particularly relevant for the detection of fast rotating pulsars that are suitable for Pulsar Timing Arrays and the detection of gravitational waves \citep{Shannon2025_SKA_SKAPTA}.

A more complete record of past surveys, including standardised reporting of pulsar parameters and detection context, would enable a more comprehensive comparison between population synthesis frameworks. Even within the snapshot and evolutionary approaches used here, different assumptions about luminosity prescriptions, beaming models, and spatial distributions can produce markedly different results (see, e.g., \citealt{xwwh22,xzx+23} for different snapshot approaches and \citealt{fk06,gmvp14,cbo20,dpm22,sn24,spmd24} for various evolutionary population synthesis frameworks).

Making full use of existing data is not only important for refining SKA yields, but also for addressing broader open questions in neutron star astrophysics. One such issue is the birth rate of different neutron star classes, which has implications for high-energy astrophysical phenomena such as GRBs, FRBs, and superluminous supernovae \citep{bwt+25,mgl+25}. As \citet{kk08} outlined, there remains a gap between the birth rates inferred for different neutron star sub-populations and the overall core-collapse supernova rate. The evolutionary population synthesis frameworks outlined above estimate birth rates of about two neutron stars per century, which is consistent with supernova rates \citep{rvc21}. However, these studies only focus on radio pulsars and do not account for other classes such as magnetars or rotating radio transients \citep{Levin2025_SKA_NSpop}. Ultimately, calibrating population synthesis models against SKA survey results will be crucial for resolving these open issues and understanding the formation of neutron stars as well as evolutionary pathways between different neutron star classes.

Our final conclusions are as follows:

\begin{itemize}
    \item SKA1 should expect to find $\sim 10,000$ slow pulsars and $\sim 800$ MSPs with AA*, and $\sim 20\%$ more if AA4 is realised or if longer surveys using SKA1-Mid are possible.
    \item The larger effective area and survey speed of SKA1-Low mean that covering extensive areas of the sky with this telescope is essential to maximise the pulsar yield. This also has the advantage of sampling the off-plane older evolved pulsar population to help us understand evolutionary questions in neutron star science.
    \item Pulsar population synthesis is a difficult endeavour with large systematic uncertainties. Nonetheless, major improvements could be made in our understanding of the neutron star population, and as an offshoot also of our survey yield modelling, with standardised reporting procedures from archival pulsar surveys. Standarised reporting will also be important for future SKA observations.
    \item With the RFI environment rapidly deteriorating, even at the SKA sites, it makes sense to perform pulsar surveys as early as possible with SKA, i.e., to start the pulsar surveys on day 1 of operations. Pulsar surveys could even commence in commissioning time. As these surveys will use some of the highest resolution data products, they are also optimised for stress testing and refining the entire observational system while additionally providing relatively quick results --- pulsars will be discovered with some regularity, as opposed to other experiments involving long-term data accumulation. This also fits in well with the finite resources available in SKA's Science Data Processor (where imaging is performed), i.e. the SKA will need to perform non-imaging observations (which take data directly from the Central Signal Processor), to enable the system to `catch up' with the flow of data to be imaged.
\end{itemize}

\section*{acknowledgements}
The authors would like to thank the referee for this paper, Duncan
Lorimer, Kaustubh Rajwade and Fabian Jankowski for useful comments
which improved the quality of the manuscript. E.~F.~K. would like to
thank Bhal Chandra Joshi, Marta Burgay and Aris Karastergiou for their
tireless work coordinating the papers for this special issue and the
ongoing work of the SKA Pulsar Science Working Group. Special thanks
go to B.~W.~Stappers and (again to) A.~Karastergiou for leading the
development of the PSS. Part of the data production, processing, and
analysis tools for this paper have been implemented and operated at
the Port d’Informaci\'o Cient\'ifica (PIC) data center. PIC is
maintained through a collaboration of the Institut de F\'isica
d’Altes Energies (IFAE) and the Centro de Investigaciones
Energ\'eticas, Medioambientales y Tecnol\'ogicas (Ciemat). V.~G. is
supported by a UKRI Future Leaders Fellowship (grant number
MR/Y018257/1). O.~A.~J.  acknowledges the support of Breakthrough
Listen, which is managed by the Breakthrough Prize Foundation, and of
the School of Physics at Trinity College Dublin. C.~P.~A. and M.~R.
are partially supported by the ERC via the Consolidator grant
``MAGNESIA'' (No.  817661), the ERC Proof of Concept "DeepSpacePULSE"
(No. 101189496), and by the program Unidad de Excelencia Mar\'ia de
Maeztu CEX2020-001058-M. C.~P.~A.’s work has been carried out within
the framework of the doctoral program in Physics at the Universitat
Autonoma de Barcelona.


\bibliography{refs.bib}
\bibliographystyle{aasjournal}
\vspace{10pt} 

\end{document}